\documentclass[%
 reprint,onecolumn,
 amsmath,amssymb,
 aip, 12pt
]{revtex4-2}
\usepackage{xcolor}
\usepackage{graphicx}
\usepackage{dcolumn}
\usepackage{bm}
\usepackage{hyperref}
\usepackage{setspace}
\usepackage{threeparttablex}
\usepackage{booktabs}
\usepackage{array}
\newcommand{\bra}[1]{\ensuremath{\left\langle#1\right|}}
\newcommand{\ket}[1]{\ensuremath{\left|#1\right\rangle}}
\newcommand{\ketbra}[2]{\ensuremath{\left|#1\right\rangle\left\langle#2\right|}}
\newcommand{\bracket}[2]{\ensuremath{\left\langle #1 \middle| #2 \right\rangle}}

\newcolumntype{P}[1]{>{\centering\arraybackslash}p{#1}}
\begin{document}

\title{The Interplay between Disorder, Local Relaxation and Collective Behaviors for an ensemble of emitters outside vs inside cavity}
\author{Zeyu Zhou}
\affiliation{Department of Chemistry, University of Pennsylvania, 231 South 34th Street, Philadelphia, Pennsylvania 19104, United States}
\author{Hsing-Ta Chen}
\affiliation{Department of Chemistry, University of Pennsylvania, 231 South 34th Street, Philadelphia, Pennsylvania 19104, United States}
\affiliation{Department of Chemistry and Biochemistry, University of Notre Dame, 251 Nieuwland Science Hall, Notre Dame, Indiana 46556, USA}
\author{Joseph E. Subotnik}
\affiliation{Department of Chemistry, University of Pennsylvania, 231 South 34th Street, Philadelphia, Pennsylvania 19104, United States}
\author{Abraham Nitzan}
\affiliation{Department of Chemistry, University of Pennsylvania, 231 South 34th Street, Philadelphia, Pennsylvania 19104, United States}
\affiliation{Department of Chemistry, Tel Aviv University, Tel Aviv 69978, Israel}

\date{\today}

\begin{abstract}
The interplay between collective optical response and molecular static and dynamic disorder is studied using simple effective Hamiltonians for an ensemble of two-level emitters inside and outside a single-mode cavity. 
We model environmental disorder by randomly modulating the molecular transition frequencies and the coupling between the emitters and the electromagnetic field. We also consider effects of intermolecular interactions and orientational disorder.
We investigate how these effects lead to new features in the steady state absorption (outside the cavity), transmission spectra (inside the cavity) and the yield of local molecular processes such as a unimolecular reaction. 
Outside the cavity, the collective behavior is manifested in the linewidth of the steady state absorption, the emission spectrum, and the local chemical yield. 
Inside the cavity, however, the collective behavior primarily determines the Rabi splitting.
Effects of intermolecular interactions under orientational disorder are also studied. 
For the most part, for all types of disorder, if we increase disorder, we find a reduction in the collective nature of the molecular response (smaller effective $N$), and therefore, the Rabi splitting contraction occurs with orientational disorder. 
Moreover, we find that static disorder is more destructive to collective behavior than dynamic disorder. 
\end{abstract}
\maketitle

\newpage

\section{Introduction}
Understanding the interplay between collective molecular optical response and molecular disorder has drawn a great deal of interest in recent years. 
Collective optical response could be seen in the time domain, for instance, Dicke superradiance emission\cite{dicke1954coherence, spano_superradiance_1989, lim_exciton_2004}, superfluorescence\cite{bonifacio1975cooperative, raino2018superfluorescence}, superabsorption\cite{raimond1982collective} have been discussed extensively in the literature, or in the frequency domain, e.g. in optical cavities where collective response is manifested in the observed Rabi splitting that characterizes the strong light-matter coupling regime in such a system. 
Near metal interface and in optical cavities, the response of an individual molecule is affected also by the confined character of the cavity as demonstrated by the Purcell effect as well as different realizations of surface enhanced spectroscopies. 
\cite{weisbuch1992observation, heinzen1987enhanced, jhe1987suppression, de1987anomalous, bayer2001inhibition, houdre1994room, gerard1998enhanced, yoshie2004vacuum, masiello2008many, spano_spectral_2010, mirsaleh2012single, spano_optical_2015, herrera_cavity_controlled_2016, herrera_absorption_2017, abid2017temperature, bisht2018collective,hertzog2019strong,  sidler2020polaritonic, smith2021exact, li2022molecular} Collective response is characterized by a non-trivial dependence of experimental observables on the number $N$ of involved molecules; for example, one finds a rate linear in $N$ for superradiance emission and a Rabi splitting $\sqrt{N}$ in the optical response of molecular optical cavities.
The dependence of these behaviors on local disorder has been the subject of several recent studies.\cite{renaud1977nonstationary, gross1982superradiance, celardo2013interplay, biella_subradiant_2013, biella2013subradiant, delga2014quantum, celardo2014cooperative, pustovit2014energy, goban2015superradiance, norcia2016superradiance, asenjo2017exponential, kirton2017suppressing, shahbazyan2000mesoscopic, sukharev2017optics, fauche2017plasmonic, ribeiro2018polariton, herrera_theory_2018, gomez2019energy, wang_quantum_2019, wang2020coherent, mattiotti2020thermal, wang_theory_2020, spano_excitonphonon_2020, wang_coherent--incoherent_2020, lee_theory_2021, wang_simple_2021, herrera2022disordered}

In this paper, we discuss the effect of different types of disorder on the collective optical response of molecular systems in and out of the cavity using two simple models: site energy disorder and coupling disorder.
We also discuss how the interplay between collective optical response (outside vs inside a single-mode cavity) and local molecular processes might affect the yield of molecular photoprocesses such as photochemical reactions.

\section{Model and Calculation Method}
\subsection{A molecular ensemble in the single exciton subspace\label{subsec: modelandpump}}
Consider an ensemble of 2-level molecules whose spatial extent is assumed for shorter than the wavelength of its resonant absorption, subjected to pumping by a near resonant classical field and further exhibiting collective emission. In addition, each molecular excited state can decay due to interaction with its local environment or a local chemical process. The time evolution of this system can be described by the following Hamiltonian:
\begin{align}
    \hat{H} =& \hat{H}_{0} + \hat{V}_\text{pump} - i\Gamma_{rad}\hat{Q}-i\Gamma_{loc}\hat{C}\label{eq:1}\\
    \hat{H}_{0} =&E_{0}\ketbra{0}{0}+ \sum_{m}E_{m}\ketbra{m}{m}
    \\
    \hat{V}_\text{pump}  =& \sum_{m}V_{0 m}\cos(\omega t)\ketbra{0}{m}+V_{m 0}\cos(\omega t)\ketbra{m}{0}\label{eq:3}
    \\
    \hat{Q} =& \sum_{m, m'}\ketbra{m}{m'}
    \\
    \label{eq:5}\hat{C} =&\sum_{m}\ketbra{m}{m}
\end{align}
In eq (\ref{eq:1})-(\ref{eq:5}), the molecular states are $\left\{\ket{0}, \ket{m}\right\}$: $\ket{0}=\prod_{j}\ket{g_j}$ is the total ground state and  $\ket{m}=\ket{x_{m}}\prod_{j\neq m}\ket{g_j}$ is a singly-excited molecular state. Note that we are working in a single-exciton basis, so that our model can describe only weak excitations where many exciton correlations can be disregarded. 
The radiative emission of the molecules is treated implicitly using the effective non-Hermitian operator $- i\Gamma_{rad}\hat{Q}$ that arises when one considers one idealized wide band radiative continuum (with density of state $\rho_{dos}$) coupled identically to all molecular emitters ($V_{j} = V_{rad}, \forall j$), leading to the purely imaginary self-energy
\begin{align}
    \Gamma_{rad}=2\pi\rho_{dos}|V_{rad}|^2
\end{align}
If we denote the wavefunction of the molecular system by $c_{m}(t)$,
then under radiative pumping of the molecules, the outgoing radiative flux at any given time $t$ is 
\begin{align}
    J_{rad}(t) = \sum_{m, m'}\Gamma_{rad}c_{m}(t)c^{*}_{m'}(t)
    \label{eq: jrad}
\end{align}
Finally, $-i\Gamma_{loc}\hat{C}$ represents decay through a local channel involving a single molecule process. In contrast to the radiative decay, this channel describes a process defined by the state of each individual molecule (and does {\em not} depend on intermolecular coherence). The outgoing \textit{local} flux at any given time $t$ is
\begin{align}
    J_{loc}(t) = \sum_{m}\Gamma_{loc}|c_{m}(t)|^2
    \label{eq: chemrxn}
\end{align}

Note that in the Hamiltonian (eq \ref{eq:1}), pumping is done by a continuous wave (CW) field at the driving frequency $\omega$. Therefore, if we plot total steady state outgoing flux ($J_{rad}+J_{loc}$) as a function of this frequency $\omega$, we obtain the steady state absorption spectrum.

\subsection{A molecular ensemble in an optical cavity\label{subsec: cavitymodel}}
To describe the same molecular ensemble inside a single-mode optical cavity, we modify the Hamiltonian (eq (\ref{eq:1})-(\ref{eq:5})) in several ways.\cite{kavokin2017microcavities}
First, the external pumping does not directly couple to the emitters, but instead to the cavity mode $\ket{v}$.
Second, the molecules do not couple directly to the far EM field, and are assumed to couple only to the cavity mode. Hence, the term $-i\Gamma_{rad}\hat{Q}$ is replaced by the coupling term (eq \ref{eq:10}) below). Finally, the radiative ($\Gamma_{cav}^{(R)}$) and nonradiative ($\Gamma_{cav}^{(NR)}$) decay rates of the cavity mode to the far field and to damping in the cavity mirrors, respectively, are taken into account. The corresponding Hamiltonian is
\begin{align}
    \hat{H} =& \hat{H}_{0} + \hat{V}_\text{cav-mol}  + \hat{V}_\text{pump} - i\Gamma_{\text{loc}}\hat{C}-i\Gamma_{\text{cav}}\ket{v}\bra{v}\label{eq: hamincav}\\
    \hat{H}_{0} =& E_{v}\ketbra{v}{v} +E_{0}\ketbra{0}{0}+ \sum_{m}E_{m}\ketbra{m}{m}
    \\
    \hat{V}_\text{cav-mol}  =& \sum_{m}V_{vm}\ketbra{v}{m} + V_{mv}\ketbra{m}{v}\label{eq:10}
    \\
    \hat{V}_\text{pump}  =& V_{0 v}\cos(\omega t)\ketbra{0}{v}+V_{v 0}\cos(\omega t)\ketbra{v}{0}
    \\
    \hat{Q} =& \sum_{m, m'}\ketbra{m}{m'}
    \label{eq: nhpart2}
\end{align}
where, $v$ is the excited state of the cavity mode (and the ground state $\ket{0}$ refers to both the cavity and the molecular systems). Similar to many previous works, we consider only a single cavity mode and in correspondence with the single-exciton model, we include only the ground and first excited states of this mode.

Here, the cavity photon leaking rate ($\Gamma_{\text{cav}} = \Gamma_{\text{cav}}^{(R)} + \Gamma_{\text{cav}}^{(NR)}$) determines the quality factor $\hbar\omega_{\text{cav}} / \Gamma_{\text{cav}}$.
In addition to the local relaxation channel (which is the same as in eq \eqref{eq: chemrxn}), the outgoing radiative flux through the cavity is
\begin{align}
    J_{cav}(t) = \Gamma_{cav}^{(R)}|c_{v}(t)|^2
    \label{eq: jcav}
\end{align}
This flux represents the combined transmission spectrum.

\subsection{Disorder and intermolecular couplings \label{section: disorder}}
Disorder can be implemented in the Hamiltonian (\ref{eq:1}) and (\ref{eq: hamincav}) by assuming that either the molecular transition energies $E_{m}$ or the molecular couplings with its radiative environment ($V_{m0}$ in eq \ref{eq:3} or $V_{mv}$ in eq \ref{eq:10}) have a random component. This random component can be constant in time (static disorder) or time-dependent (dynamic disorder)

\subsubsection{Energy disorder}
If we focus first on molecular transition energy disorder, the static disorder limit is described by including a random component in the individual molecular frequencies ($\omega_{m} = E_{m}/\hbar$)
\begin{equation}
\omega_{m}=\omega_{0}+\Omega_{m}.
\end{equation} 
In the calculations reported below, the random component $\Omega_{m}$ is sampled from a Gaussian distribution
\begin{equation}
P[\Omega_{m}]=\frac{1}{\sqrt{2\pi\sigma^{2}}}\exp{\left(-\Omega_{m}^{2}/2\sigma^{2}\right)}
\end{equation}
where different molecules are assumed to be uncorrelated, $\langle\Omega_{m}\Omega_{m'}\rangle=\delta_{mm'}\sigma^{2}$ and here the variance $\sigma^{2}$ indicates the disorder strength. This static disorder model may be viewed as the static limit of Kubo’s stochastic modulation model where
\begin{equation}
    \omega_{m}=\omega_{0}+\Omega_{m}(t)
\end{equation}
and $\Omega_{m}(t)$ are stochastic random variables that change over time. We choose $\Omega_{m}(t)$ to be a Gaussian stochastic variable that satisfies $\langle\Omega_{m}(t)\rangle=0$ and
\begin{equation}
\langle\Omega_{m}(t_{1})\Omega_{m'}(t_{2})\rangle=\delta_{mm'}\sigma^{2}e^{-|t_{1}-t_{2}|/\tau_{c}}
\end{equation}
where $\delta_{mm'}$ is a Kronecker delta function. 
The Gaussian stochastic variables are characterized by the amplitude $\sigma = \langle\delta\Omega^2\rangle^{1/2}$ and correlation lifetime $\tau_{c}$ of local energy fluctuations. 
For a detailed algorithm to generate such a stochastic process, see Appendix \ref{apdx: ouprocess}.
\subsubsection{Orientational Disorder \label{subsubsec: orientdisordertheory}}
\begin{figure}[!ht]
\includegraphics[width=10cm]{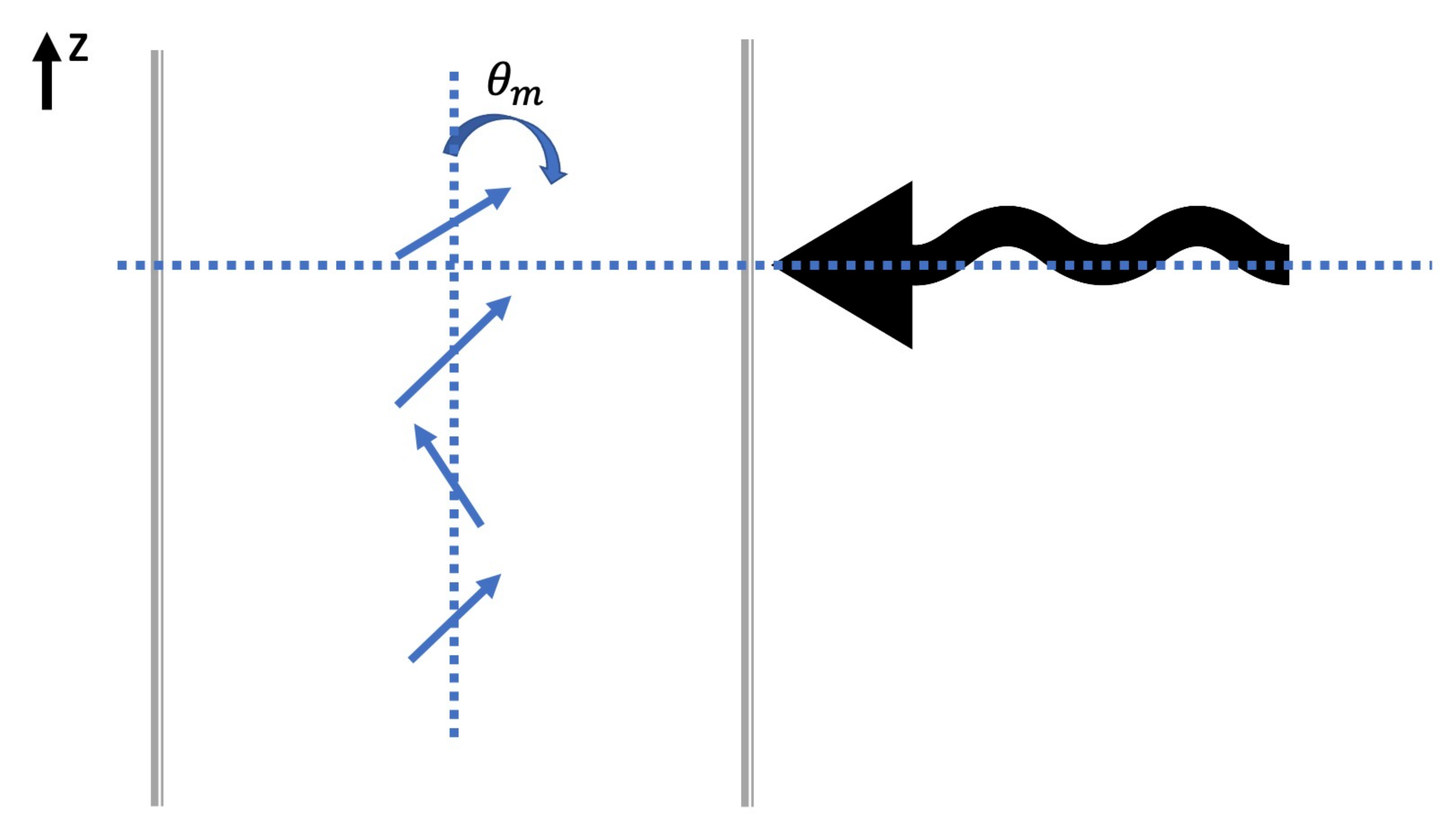}
\caption{Schematic diagram of the molecular emitters (blue arrows) confined in an optical cavity (grey walls) and pumped by the incoming CW field (black line). The field of the cavity mode is assumed to be in the $z$ direction. The transition dipole moments of the emitters have different angles $\theta_{m}$ with respect to the $z$ axis. }
\label{fig: systemdemo}
\end{figure}

The most obvious source of coupling disorder is the inherent orientational disorder that characterizes most molecular liquids, and here we focus on this aspect of our systems.
The simplest way to model such disorder is to assign an angle $\theta_{m}$ for each emitter $m$, where $\theta_m$ is oriented with respect to the cavity mode (see Fig.~\ref{fig: systemdemo}). 
We assume that the field of the cavity mode that may be excited by the external pumping is polarized in the $z$ direction and denote $\theta_{m}$ as the angle between the $z$ axis and the molecular transition dipole direction. 
This disorder leads to three modifications of the model Hamiltonian: First, the cavity-molecule coupling becomes
\begin{equation} 
\hat{V}_\text{cav-mol}  = \sum_{m}V_{vm}\cos(\theta_{m})\ketbra{v}{m}+ V_{mv}\cos(\theta_{m})\ketbra{m}{v}
\label{eq: staticangulardisordercos}
\end{equation}
Similarly, for the case outside the cavity, the couplings to the external driving EM field also depend on the emitters orientation
\begin{equation} 
\hat{V}_\text{pump}  = \sum_{m}V_{0 m}\cos(\theta_{m})\cos(\omega t)\ketbra{0}{m}+V_{m 0}\cos(\theta_{m})\cos(\omega t)\ketbra{m}{0}
\end{equation}
Also, outside the cavity, the elements of the non-Hermitian $\hat{Q}$ matrix depend on the relative angle $\cos(\theta_{m} -\theta_{m'} )$\cite{akram2000decoherence}
\begin{equation} 
\hat{Q} = \sum_{m, m'}\cos(\theta_{m} -\theta_{m'} )\ketbra{m}{m'}
\end{equation}

The disorder of $\{\theta_m\}$ can be either static or dynamic as described in section~\ref{section: disorder}.  
At room temperature, the characteristic timescale of the rotational motion in solution is of order $10ps$, which implies that in most situations including light-matter strong coupling regime in molecular system, orientational disorder may be assumed static. Nevertheless, for completeness, we will consider below a full range of dynamic disorder.
In the static case, we choose a time-independent $\{\theta_m\}$ according to random orientation angles sampled from a Gaussian distribution
\begin{equation}
P[\theta_{m}]=\frac{1}{\sqrt{2\pi\sigma^{2}}}\exp{\left(-\theta_{m}^{2}/2\sigma^{2}\right)}
\label{eq: orientdisor}
\end{equation}
This again is the limit $\tau_{c}\rightarrow \infty$ of a process where orientations are dynamically modulated according to $\langle\theta_{m}(t)\rangle=0$ and 
\begin{equation}
\langle\theta_{m}(t_{1})\theta_{m'}(t_{2})\rangle=\delta_{mm'}\sigma^{2}e^{-|t_{1}-t_{2}|/\tau_{c}}
\label{eq: orientdisorcorrela}
\end{equation}
\subsubsection{Orientational disorder with intermolecular coupling}
To complete our investigation of linear response of molecules outside and inside optical cavities as expressed by the absorption and transmission signals, we will also consider models with intermolecular couplings (dipole-dipole couplings), again focusing on static and dynamic orientational disorder. The intermolecular couplings is modeled by a dipole-dipole coupling: 
\begin{align}
    \hat{V}_{\text{dip}} =&\sum_{m=k+1}^{N}\sum_{k=1}^{n}W_{k}f(\theta_{m}, \theta_{m-k}) \ketbra{m}{m-k}+\sum_{m=k+1}^{N}\sum_{k=1}^{n}f(\theta_{m-k}, \theta_{m})W_{k} \ketbra{m-k}{m}
    \label{eq: intermolcoupling}
\end{align}
Here, $W_{k} = W / k^3$ and in the calculations reported below we have used $W = 0.1$,\footnote{The choice of intermolecular coupling is $0.1$ in unit of Rabi splitting. Rabi splitting is approximately on the order of $10 meV$, and if the distance between dipoles are $2 nm$, the dipole moments are on the order of $10 Debye$.} unless otherwise specified. In principle, this Hamiltonian describes a chain of emitters and the interaction between them decays as one over their distance cubed, $1/r^3$, as sketched in Fig~\ref{fig: systemdemo}.

The function $f(\theta_{m}, \theta_{m-k})$ in eq \ref{eq: intermolcoupling} reflects the angular dependence of the dipole-dipole coupling between emitters and is given by
\begin{align}
    f(\theta_{m}, \theta_{m-k}) = \cos(\theta_{m}- \theta_{m-k}) - 3\cos\theta_{m}\cos\theta_{m-k}
    \label{eq: dpdpformula}
\end{align}
Again, these angles are defined to be relative to the polarization of the incoming driving electric field.
\subsection{The Steady State flux}
In order to simulate cavity-molecular dynamics, we propagate the Schr\"odinger equation for the Hamiltonians above under steady state boundary conditions. 
The wavefunction for the driven, open quantum system can be written as  $\ket{\Psi}=c_{0}\ket{0}+c_{v}\ket{v}+\sum_{m}c_{m}\ket{m}$.
This implies propagating the Schr\"odinger equation $i\hbar\frac{d}{dt}\ket{\Psi}=\hat{H} \ket{\Psi}$ under the Hamiltonians of eqs \ref{eq:1} and \ref{eq: hamincav} keeping $|c_{0}|^2=1$. The long-time evolution yields the steady state forms of the coefficients $c_{m}(t)$ and $c_{v}(t)$ that are used to evaluate the steady state fluxes $J_{rad}$ (eq \ref{eq: jrad}),
$J_{loc}$ (eq \ref{eq: chemrxn}), and 
$J_{cav}$ (eq \ref{eq: jcav}). Again, the total steady state flux $J_{rad}+J_{loc}$ as a function of the incoming driving frequency yields the steady state absorption spectrum for emitters outside optical cavities; while inside the cavity, the steady state flux $J_{cav}$ gives reflection/transimission spectrum and $J_{loc}$ represents the local relaxation flux. 

With this single-exciton model, there are many parameters we can adjust to account for different physical conditions. We will discuss these different possibilities in the following sections.

\section{Results}
\subsection{Energy disorder\label{subsubsec: resnocavsddd}}
\begin{figure}[!ht]
\includegraphics[width=15cm]{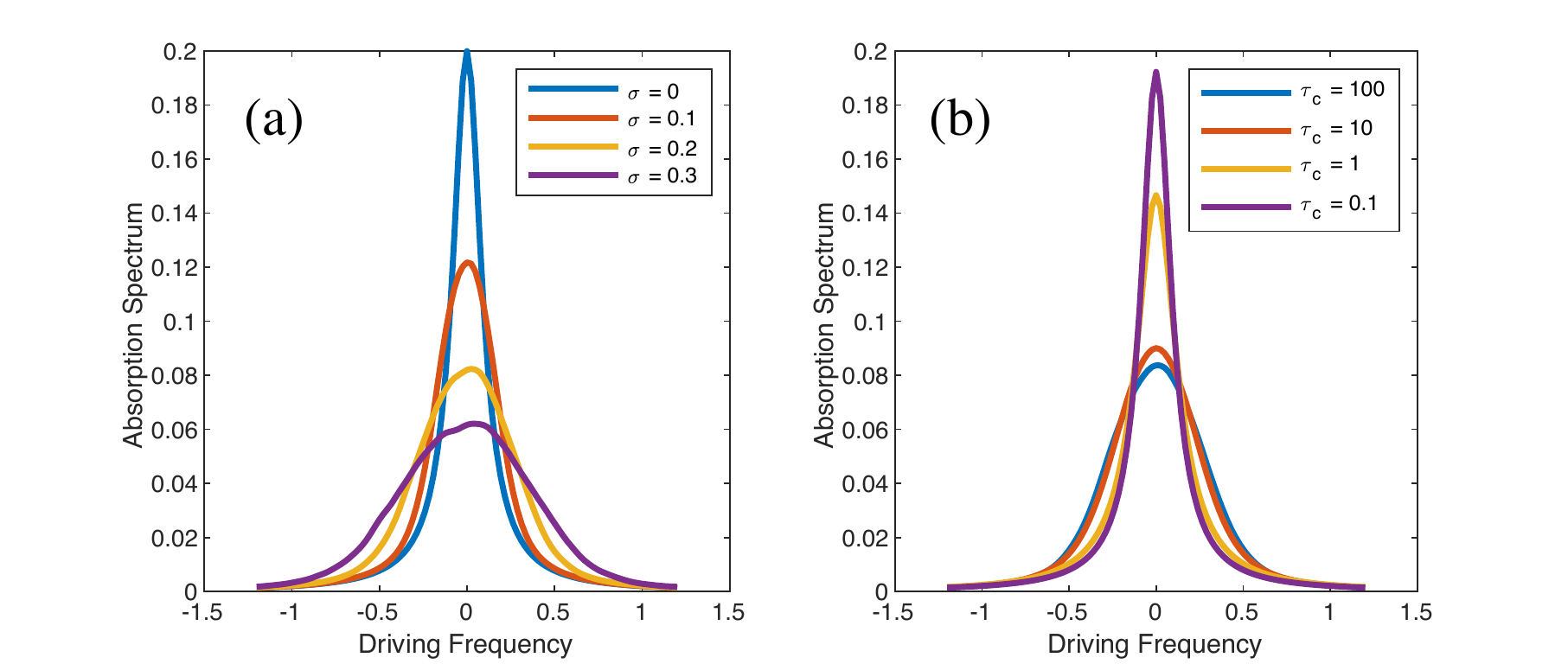}
\caption{The steady state absorption spectrum for a molecular ensemble outside the cavity with (a) different static energetic disorder strengths and (b) different correlation time $\tau_{c}$ but constant disorder strength $\sigma = 0.2$. The molecular ensemble includes $N=40$ molecules. If there is no disorder (panel (a), $\sigma = 0$, blue line), the linewidth of the peak is $2N\Gamma_{rad} + 2\Gamma_{loc}$. As the static disorder strength increases, we obtain a broader inhomogeneous peak. As the correlation time decreases under fixed disorder strength (panel (b), $\sigma = 0.2$), the peak gets narrower, reflecting motional narrowing.
As shown in eqs \ref{eq: ddvoigt} and \ref{eq: sdvoigt}, as well as Appendix \ref{apdx: voigt_theory}, these features can be explained analytically.}
\label{fig: outsidecavitysddd}
\end{figure}

In Fig~\ref{fig: outsidecavitysddd}, we plot the absorption spectrum for an ensemble of emitters outside the cavity. 
In the limit of zero disorder, the linewidth of the absorption spectrum is associated with the decay rate for the radiative and local non-radiative channels.
For the static disorder results (a), as the disorder strength $\sigma$ increases, the absorption lineshape become lower and broader.
This feature reflects inhomogeneous broadening and the static uncertainty in the molecular excitation energy. 
Under dynamic disorder (b), as the correlation time $\tau_{c}$ decreases, the absorption linewidth is smaller, which reflects motional narrowing.
We can also predict the absorption lineshape in Fig~\ref{fig: outsidecavitysddd} by performing a Fourier transform of $\phi(t)$\footnote{This formula is by modifying equation (7.104) in ref. \citenum{nitzan2006chemical} by taking both decay rates $N\Gamma_{rad}+\Gamma_{loc}$ into account according to Voigt theorem.}
\begin{align}
    \phi(t) = \exp[-\sigma^2\tau_{c}^2(\frac{t}{\tau_{c}}-1+e^{-\frac{t}{\tau_{c}}})-(N\Gamma_{rad}+\Gamma_{loc})t].
    \label{eq: ddvoigt}
\end{align}
For the case of static disorder ($\tau_{c}\rightarrow\infty$), we can simplify the formula by expanding the exponential of $t/\tau_{c}$
\begin{align}
    \phi(t) \rightarrow \exp(-\sigma^2t^2/2-(N\Gamma_{rad}+\Gamma_{loc})t)
    \label{eq: sdvoigt}
\end{align}
which, after a Fourier transform, yields a convolution between the gaussian associated with static disorder and the lorentzian associated with the radiative and non-radiative decay rates. For more details and a comparison of analytic versus numerical results, see Appendix \ref{apdx: voigt_theory}.

\begin{figure}[!ht]
\includegraphics[width=15cm]{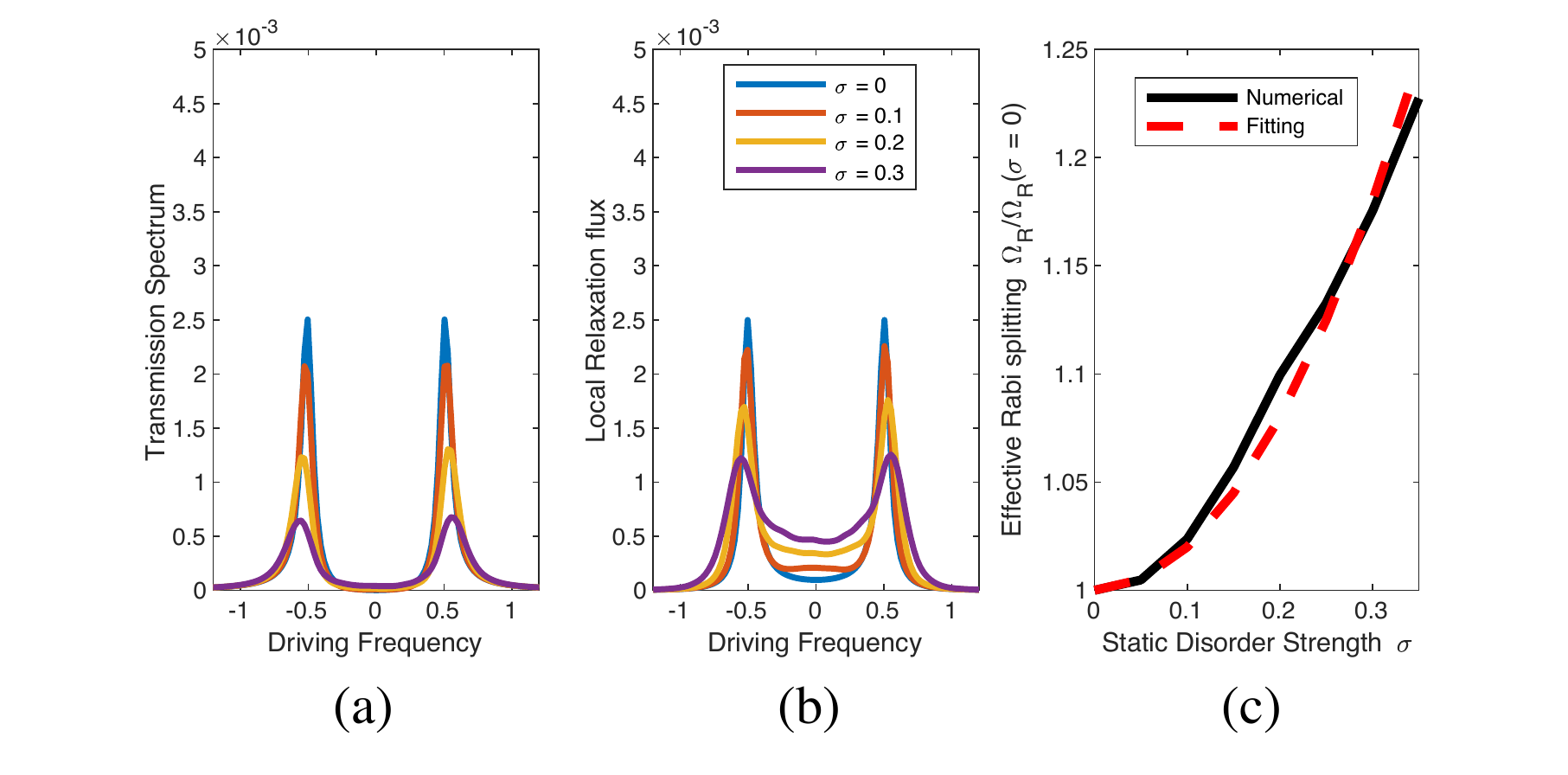}
\caption{Steady state (a) transmission spectrum  and (b) local relaxation flux spectra, as calculated for molecular ensembles with different static disorder strengths $\sigma$ {\em interacting with a single-mode cavity}. The molecular ensemble and the cavity mode form two polariton peaks in both figures. As the amplitude of static disorder increases, the split peaks become broadened; however, the baseline in between the two peaks (around driving frequency $\omega_{d} = 0$) grows larger only for the local signal. This growth arises because the dark modes contribute only to the local relaxation flux but not to the transmission spectrum. Lastly, note that as the energetic static disorder increases, the effective Rabi splitting increases. This increase can be captured approximately by perturbation theory (eq \ref{eq: rabisplittingvssigma}).}
\label{fig: incavsdtranschem}
\end{figure}

\begin{figure}[!ht]
\includegraphics[width=15cm]{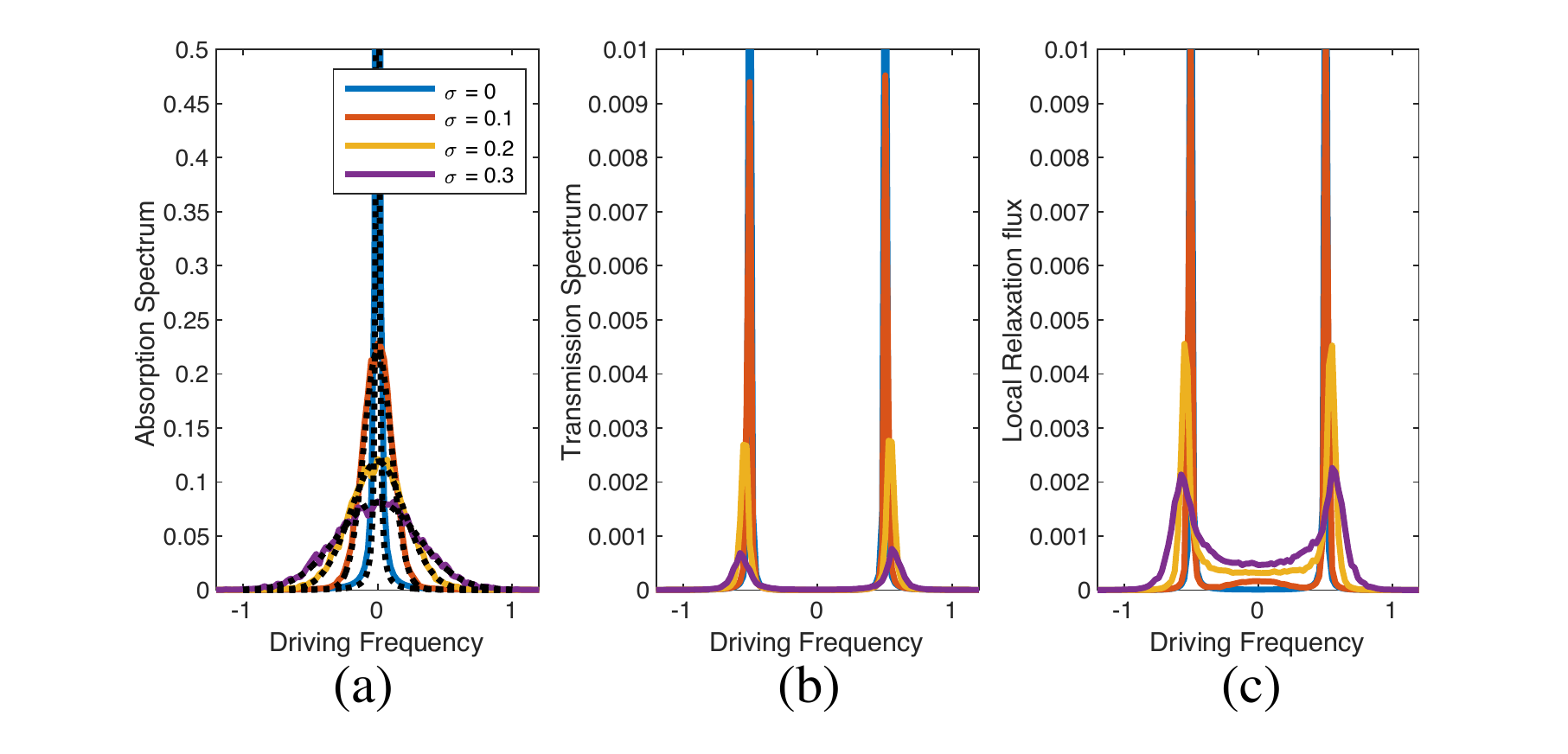}
\caption{Steady state (a) absorption spectra for  outside the cavity, (b) transmission and (c) local relaxation flux spectra {\em interacting with a single-mode cavity}, as calculated for molecular ensembles with different static disorder strengths $\sigma$.   The black dotted lines in panel (a) are analytical results calculated in Appendix \ref{apdx: voigt_theory}.  
Here, the contributions to homogeneous broadening, namely the non-Hermitian parts of the Hamiltonian matrices are taken $\Gamma_{rad}=0.005/N, \Gamma_{cav}=0.005, \Gamma_{loc} = 0.005$, which are $10$ times smaller than in Fig \ref{fig: incavsdtranschem}, so (except from the case without disorder) the linewidth outside the cavity is dominated by the inhomogeneous broadening.
In agreement with earlier observations\cite{houdre1996vacuum}, the inhomogeneous broadening observed outside the cavity is not manifested in the linewidth of the polariton peaks inside the cavity which are therefore much narrower as seen in panel (b) and (c). }
\label{fig: sdbroadening}
\end{figure}

Next, consider the case where the same ensemble of emitters interacts with a single-mode cavity under steady state pumping. As discussed in section \ref{subsec: cavitymodel}, we calculate two observables: (1) the steady state flux $J_{cav}$ through the cavity leakage channel (which gives the combined transmission spectrum) and the steady state flux $J_{loc}$ through the local relaxation channel of individual emitters.
The results in Fig. \ref{fig: incavsdtranschem} show that static disorder leads to an inhomogeneous broadening of the two polariton peaks that are present in both of the relevant channels.
Interestingly, Fig \ref{fig: sdbroadening}, in which we suppress the homogeneous contributions (i.e. when non-Hermitian parts $\Gamma_{rad}$, $\Gamma_{loc}$, $\Gamma_{cav}$ are small) to the lineshape so that the width is dominated by the inhomogeneous contribution, shows that this inhomogeneous broadening is much weaker inside a cavity relative to outside a cavity  . This effect has already been seen in reference \citenum{houdre1996vacuum}.

Returning to Fig. \ref{fig: incavsdtranschem}, we note that, for increasing $\sigma$, the baseline between the two polariton peaks (around driving frequency $\omega_{d} = 0$) for the local relaxation channel becomes larger, whereas in the transmission spectrum, the baseline around driving frequency $\omega_{d} = 0$ remains approximately zero.
This different behavior arises because the dark modes cannot contribute to the transmission spectrum but they do contribute to the local relaxation channel.
Another observation in Fig. \ref{fig: incavsdtranschem} is the increase of the effective Rabi splitting in both figures (a) and (b), as the energetic static disorder strength increases. This increase can be captured approximately by the following formula
\begin{equation}
    \Omega_{R}(\sigma) / \Omega_{R}(\sigma=0)  = 1 + 2\sigma^2
    \label{eq: rabisplittingvssigma}
\end{equation}
One can obtain such a formula by using second-order perturbation theory for the polariton states. For a simple derivation, see Appendix \ref{apdx: sdfitting}.

\begin{figure}[!ht]
\includegraphics[width=15cm]{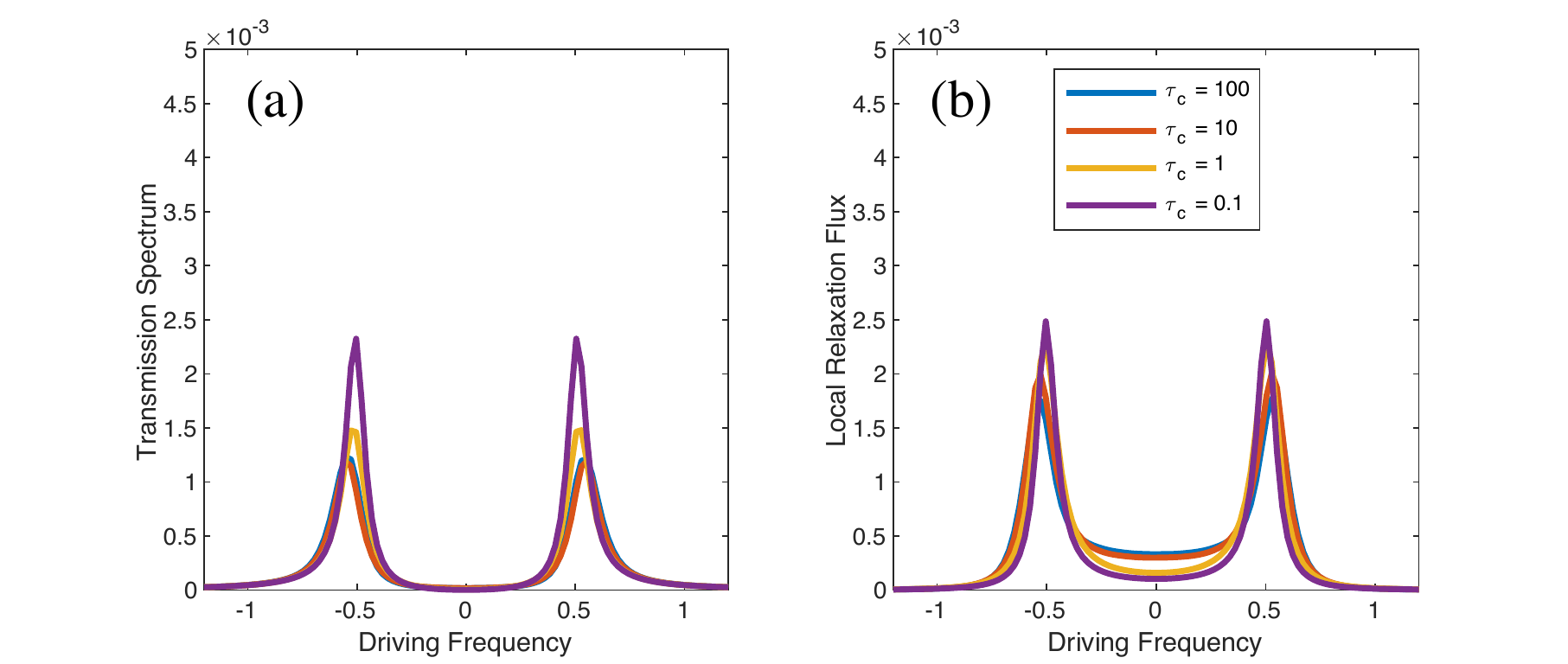}
\caption{Steady state (a) transmission spectra and (b) local relaxation for the molecular ensemble {\em inside a single-mode cavity} with dynamic disorder. We fix the disorder strength ($\sigma = 0.2$) and scan the correlation time $\tau_{c} = 100, 10, 1, 0.1$. The molecular ensemble and the cavity mode forms two polariton peaks in both figures. Note that the long correlation limit (blue lines) recovers the results with static disorder in Fig. ~\ref{fig: incavsdtranschem} (Yellow lines). Note also that similar to Fig. \ref{fig: incavsdtranschem}, the dark modes contribute only to the local relaxation channel. As we decrease $\tau_{c}$, the two peaks become narrower in both figures, and the dark modes contribute less to the local relaxation channel.}
\label{fig: incavddtranschem}
\end{figure}

Lastly, for the case with dynamic disorder, as shown in Fig. \ref{fig: incavddtranschem}, as one might expect, we find motional narrowing for both channels as the correlation time $\tau_{c}$ decreases (similar to Fig. \ref{fig: outsidecavitysddd}). Moreover, we observe the same behavior as in Fig. \ref{fig: incavsdtranschem}: in between the two polariton peaks (around driving frequency $\omega_{d} = 0$), the presence of dark modes leads to a finite steady state flux for the local relaxation channel.
\subsection{Orientational disorder\label{subsec: orientationaldisorder}}
\begin{figure}[!ht]
\includegraphics[width=15cm]{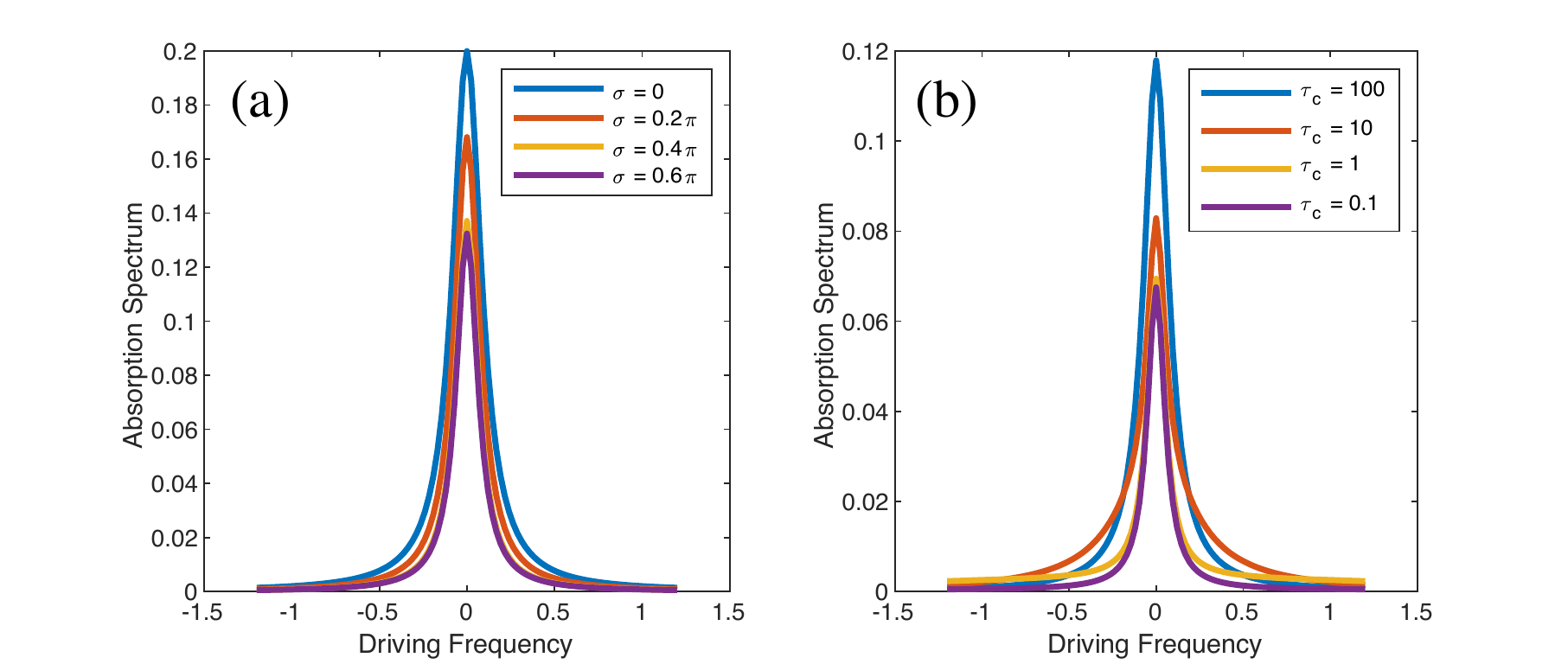}
\caption{Steady state absorption spectra for the ensemble of emitters outside the cavity. (a) static angular disorder; (b) dynamic angular disorder. As shown in the static disorder figure, when the disorder strength $\sigma$ increases, the absorption peaks decrease. More importantly, the linewidth decreases because the emitters act less collectively. For the case of dynamic disorder, similar to Fig. \ref{fig: outsidecavitysddd}, we observe motional narrowing.}
\label{fig: outcavsdddangular}
\end{figure}

In this subsection, we present results for angular disorder as defined in section \ref{subsubsec: orientdisordertheory}.
Note that the way disorder is defined in this case implies the average contribution is nonzero ($\langle\cos \theta_{m}\rangle\neq 0$), and therefore, some of the results shown below reflect this situation. 
As shown in Fig~\ref{fig: outcavsdddangular}, for the static angular disorder case, as the disorder strength $\sigma$ increases, both the height and the linewidth of the absorption peak decrease (see eq \ref{eq: orientdisor}). This decrease arises  because each dipole in the ensemble of emitters couples to the driving field differently, resulting in less collective behavior. For the dynamic angular disorder case, in agreement with the results in section \ref{subsubsec: resnocavsddd},  as we decrease the correlation time of the angular disorder, we find motional narrowing. 

\begin{figure}[!ht]
\includegraphics[width=15cm]{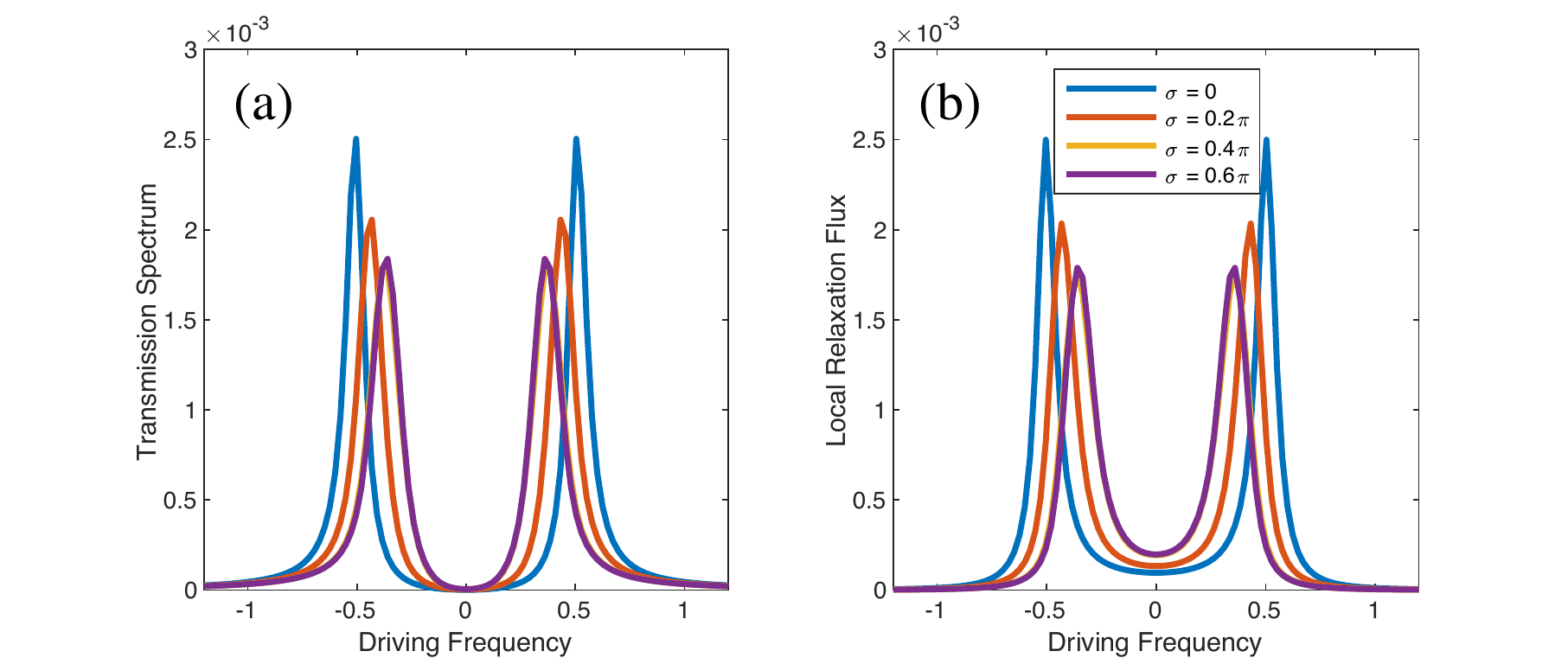}
\caption{Steady state transmission spectrum and local relaxation flux for different static angular disorder strength ($\sigma = 0, 0.2\pi, 0.4\pi, 0.6\pi$) inside a cavity. As shown in the figures, when the disorder strength $\sigma$ increases, the absorption peak height decreases and the dark modes contribute only to the local relaxation flux but not to the transmission spectrum  (similar to Fig. \ref{fig: incavsdtranschem}). However, unlike Fig. \ref{fig: incavsdtranschem}, disorder in the angular momentum leads to a decrease in the Rabi splitting. }
\label{fig: incavsdtranschemangular}
\end{figure}

Next, we consider this molecular ensemble with angular disorder inside the cavity. The results are shown in Fig. \ref{fig: incavsdtranschemangular}. 
Unlike the results in Fig. \ref{fig: incavsdtranschem}, (where we investigated disorder in the excitation energies), here the Rabi splitting consistently decreases as we increase the static angular disorder strength and lose collectivity of the response. Clearly, when interpreting the spectra from cavity, one must consider geometry and not just the energy levels of the molecules in the cavity.  In principle, in a cavity, one might expect to have both energy and angular disorder and so, as the temperature rises, the Rabi splitting should decrease. 
Unfortunately, it remains unclear which effect dominates. Note that the increase of Rabi splitting with increasing static energy disorder does not suggest that more emitters are behaving collectively. As shown both in Fig. \ref{fig: incavsdtranschem} and Fig. \ref{fig: incavsdtranschemangular}, the maximal radiative steady state flux decreases as static disorder strength $\sigma$ increases. Lastly,  as in Fig. \ref{fig: incavsdtranschem}, the dark modes contribute to the rise of baseline between the two polariton peaks only for the local relaxation channel.

\begin{figure}[!ht]
\includegraphics[width=15cm]{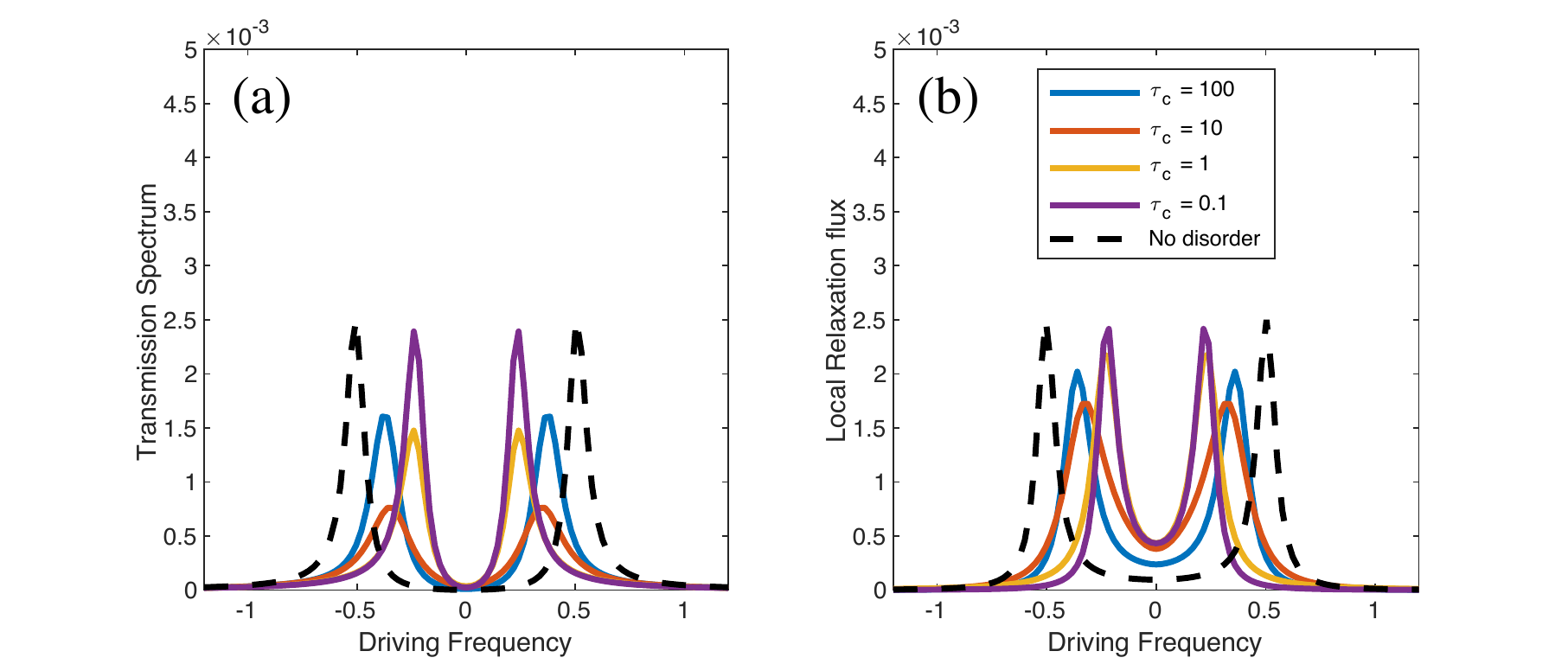}
\caption{Steady state absorption spectra for dynamic angular disorder. We fix the angular disorder strength ($\sigma = 0.4\pi$) and scan over correlation time $\tau_{c} = 100, 10, 1, 0.1$. We also plot the result with no disorder for reference. As shown in both figures, when the correlation time decreases, the collective effect diminishes, leading to a smaller Rabi splitting. Similar to Fig. \ref{fig: incavsdtranschemangular}, the dark modes contribute only to the local relaxation signals.}
\label{fig: incavddtranschemangle}
\end{figure}

The results for the dynamic angular disorder case are slightly more non-intuitive. In Fig. \ref{fig: incavddtranschemangle}, we fix the disorder strength ($\sigma=0.4\pi$) and calculate transmission and local relaxation fluxes for different correlation time $\tau_c$. In the long correlation time limit (blue line), we recover the static disorder result. Two features are worth noting in Fig. \ref{fig: incavddtranschemangle}.  First, as  $\tau_c$ becomes smaller, we find that, as expected, the polaritonic spectra undergo motional narrowing. However, reducing the correlation time is not equivalent to reducing disorder insofar as the fact that the Rabi splitting in Fig. \ref{fig: incavddtranschemangle} decreases as $\tau_c \rightarrow 0$ (whereas the Rabi splitting increases as $\sigma\rightarrow 0$ in Fig. \ref{fig: incavsdtranschemangular}).
For a simple explanation of this Rabi splitting contraction, consider the two limiting scenarios,  (i) the static limit ($\tau_c = \infty$) and (ii) the no correlation limit ($\tau_c = 0$). For the static limit ($\tau_c = \infty$), the averaged Rabi splitting is calculated by estimating the eigenvalues $E_{LP/UP}$.
\begin{align}
    E_{LP/UP} = \sqrt{N|V_{mv}|^{2}\langle\cos^2(\theta_{m})\rangle}
    \label{eq: 29}
\end{align}
Here, $\langle\rangle$ stands for ensemble average. For the no correlation limit ($\tau_c = 0$), a good estimate is
\begin{align}
    E_{LP/UP} = \sqrt{N|V_{mv}|^{2}}\langle\cos(\theta_{m})\rangle
    \label{eq: 30}
\end{align}
However, we are unable to predict what Rabi splittings are for finite $\tau_{c}$. 
For detailed derivations of these expressions and an analysis of how they match up with numerical results, see Appendix \ref{apdx: saddadtheory}. Note that in both limits, the Rabi splittings are smaller than the case with no disorder, reflecting the fact that the average coupling in this model is smaller than $V_{mv}$.

Another non-intuitive result in Fig. \ref{fig: incavddtranschemangle} pertains to the height of the peaks themselves. Here, we see that the height behaves in a non-monotonic fashion as a  function of $\tau_c$; starting from static disorder and decreasing $\tau_c$, we find that the peak height and integral decreases and then increases. Qualitatively, we believe this non-monotonic behavior can be understood by recognizing that with moderate correlation times, introducing disorder and slowly distorting the coherences removes the possibility of $(i)$ seeing individual configurations, each absorbing for a long time or $(ii)$ seeing many configurations averaged to one static configuration. As a result, there is less absorption at intermediate correlation times. Pursuing a more rigorous analysis of this effect will certainly be an important research direction in the future. 
Finally, just as in Fig. \ref{fig: incavsdtranschemangular}, the dark modes contribute to the local relaxation,  and as the correlation time decreases, the Rabi splitting decreases.

\subsection{Intermolecular coupling and orientational disorder}
\begin{figure}
\includegraphics[width=15cm]{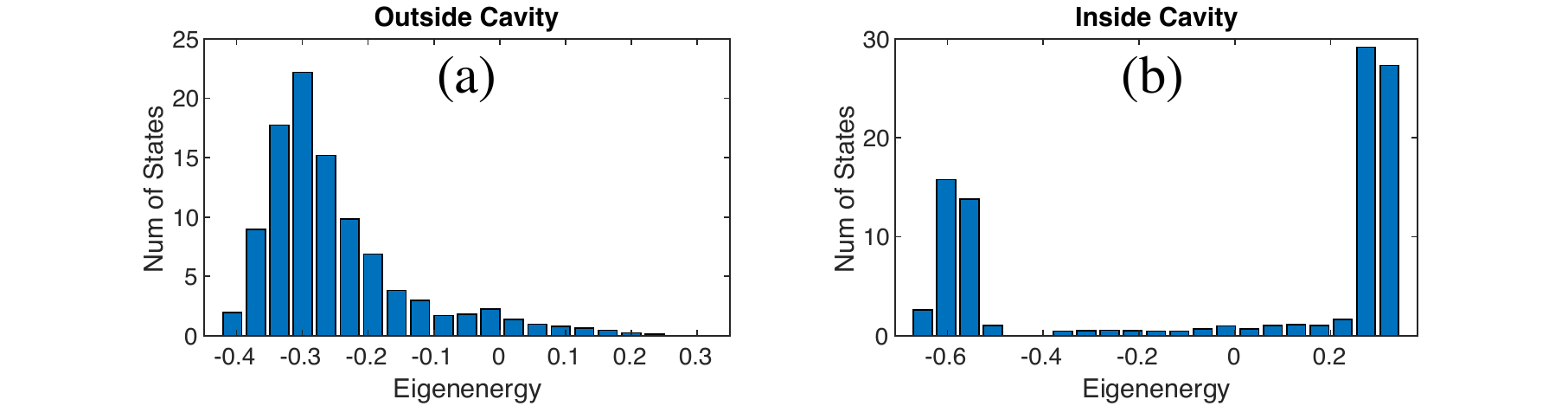}
\caption{Histogram  of the brightness ($w_j$ in eq \ref{eq: brightness} and $w_{j}^{\text{cav}}$ in eq \ref{eq: brightnessincav}) for the Hamiltonian eigenstates that arise with intermolecular coupling (see eqs \ref{eq: intermolcoupling} and \ref{eq: dpdpformula})
as averaged over realizations of $\theta_{m}$ following the distribution in eqs \ref{eq: orientdisor} and \ref{eq: orientdisorcorrela}. These brightness factors are relevant to pumping the emitters outside/inside the cavity. We consider that the dipoles are fully aligned in the limit of zero disorder and thus, intermolecular couplings are negative (eq \ref{eq: dpdpformula} equals $-2$). The eigenvalue of the bright state (symmetric superposition state) becomes negative. Hence, for the case outside the cavity, the brightest states have energy $E\approx-0.3 \ a.u.$, which is lower than the energy of the independent emitters (which is set to be $0$); for the case inside the cavity, the two brightest states yield lower and upper polariton peaks at around $E = -0.6, 0.3 \ a.u.$ respectively. The upper polariton peak is brighter than the lower polariton because the former has an energy eigenvalue ($0.3 \ a.u.$) closer to the energy of the bare cavity mode ($0 \ a.u.$). Again, the center of the two polariton peaks ($(E_{LP} + E_{UP})/2\approx-0.15 \ a.u.$) is lower than the energy of the independent emitters/bare cavity mode.}
\label{fig: densityofstatesfordifferentprobeangles}
\end{figure}

Finally, let us address the presence of how intermolecular coupling affects orientational disorder.
Let $\ket{B}$ be the bright state of the system outside the cavity
\begin{align}
    \ket{B} &= \frac{1}{\sum_{j}\cos^{2}\theta_{j}}(\cos\theta_{1}, \cos\theta_{2}, ...)^{T}.
\end{align}
Before presenting the absorption spectra, we start by showing the density of states of the system outside/inside a cavity, weighted by the brightness of the states in Fig~\ref{fig: densityofstatesfordifferentprobeangles}. 
For the case outside the cavity, we define the brightness $W_{j}$ of each eigenstate $\phi_{j}$ of $\hat{H}_{sub}$ (keeping only the real part of $m$ and $v$ elements, see eqs \ref{eq: subblockham}-\ref{eq: subblockham3} in Appendix \ref{apdx: sdfitting}) to be:
\begin{align}
    w_{j} &= |\bracket{B}{\phi_{j}}|^2
    \label{eq: brightness}
\end{align}
For the case inside the cavity, the bright mode must contain the cavity mode and thus the following definition makes more sense:
\begin{align}
    w_{j}^{\text{cav}} &= |\bracket{v}{\phi_{j}}|^2
    \label{eq: brightnessincav}
\end{align}
where $\ket{v}$ is the bare cavity state (see eqs \ref{eq: hamincav}-\ref{eq: nhpart2}). 
As shown in Fig~\ref{fig: densityofstatesfordifferentprobeangles}, for the case outside the cavity, the brightest eigenstates have negative energy eigenvalues (relative to the emitters energy). For the case inside the cavity, there are two regions of bright states corresponding to the two polariton peaks; the center of the two peaks is shifted towards the negative direction, just as for the case outside the cavity. The upper polariton eigenstates are brighter than the lower polariton because they are closer to the bare cavity mode energy.
\begin{figure}
\includegraphics[width=15cm]{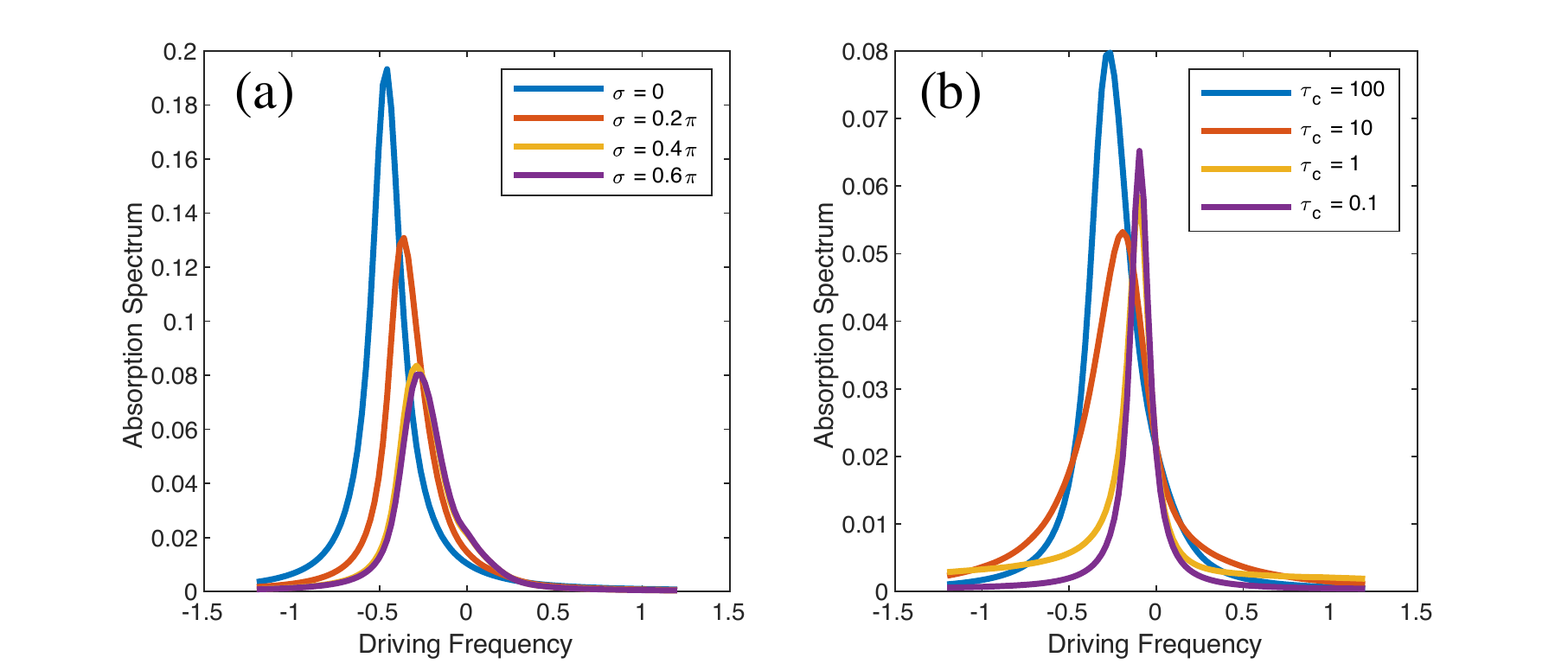}
\caption{Steady state absorption for emitters outside cavity with intermolecular couplings and static/dynamic orientational disorder. (a) static orientational disorder; (b) dynamic orientational disorder. For the static disorder case, the absorption peaks height decreases as the disorder strength increases. As predicted in Fig. \ref{fig: densityofstatesfordifferentprobeangles}, intermolecular coupling reduces the energy of the system and the peaks appear to have energy lower than that of the individual emitters. For the dynamic disorder case, the peak gets narrower as we decrease the correlation time.}
\label{fig: outcavsdddangulardip}
\end{figure}
\begin{figure}
\includegraphics[width=15cm]{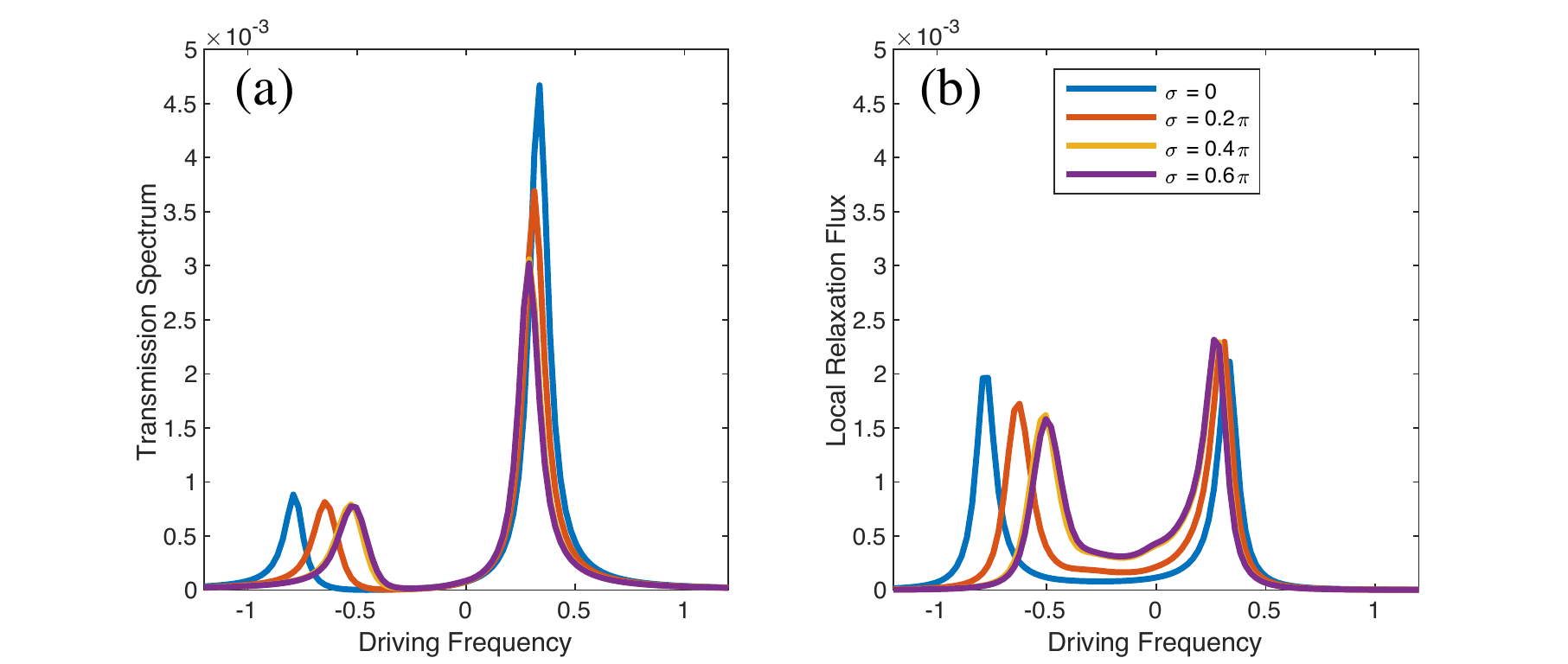}
\caption{Steady state (a) transmission and (b)local relaxation for an ensemble of emitters inside a single-mode cavity with intermolecular couplings for different static orientational disorder strength. As predicted in Fig. \ref{fig: densityofstatesfordifferentprobeangles}, the upper polariton contributes more strongly than does the lower polariton peak to the transmission spectrum. Intermolecular couplings stabilize the system and thus the center of the two polariton peaks corresponds to an energy lower than that of an individual emitter/bare cavity mode. As the disorder strength increases, the Rabi splitting decreases as the collectiveness is destroyed by the disorder. The difference in height between the upper and lower polariton peaks is small for the local relaxation channel because the signals are not dependent on how strong is the cavity mode contribution to the polariton (but rather are dictated by the dynamics of the individual emitters).}
\label{fig: incavdipsd_transvschem}
\end{figure}
\begin{figure}
\includegraphics[width=15cm]{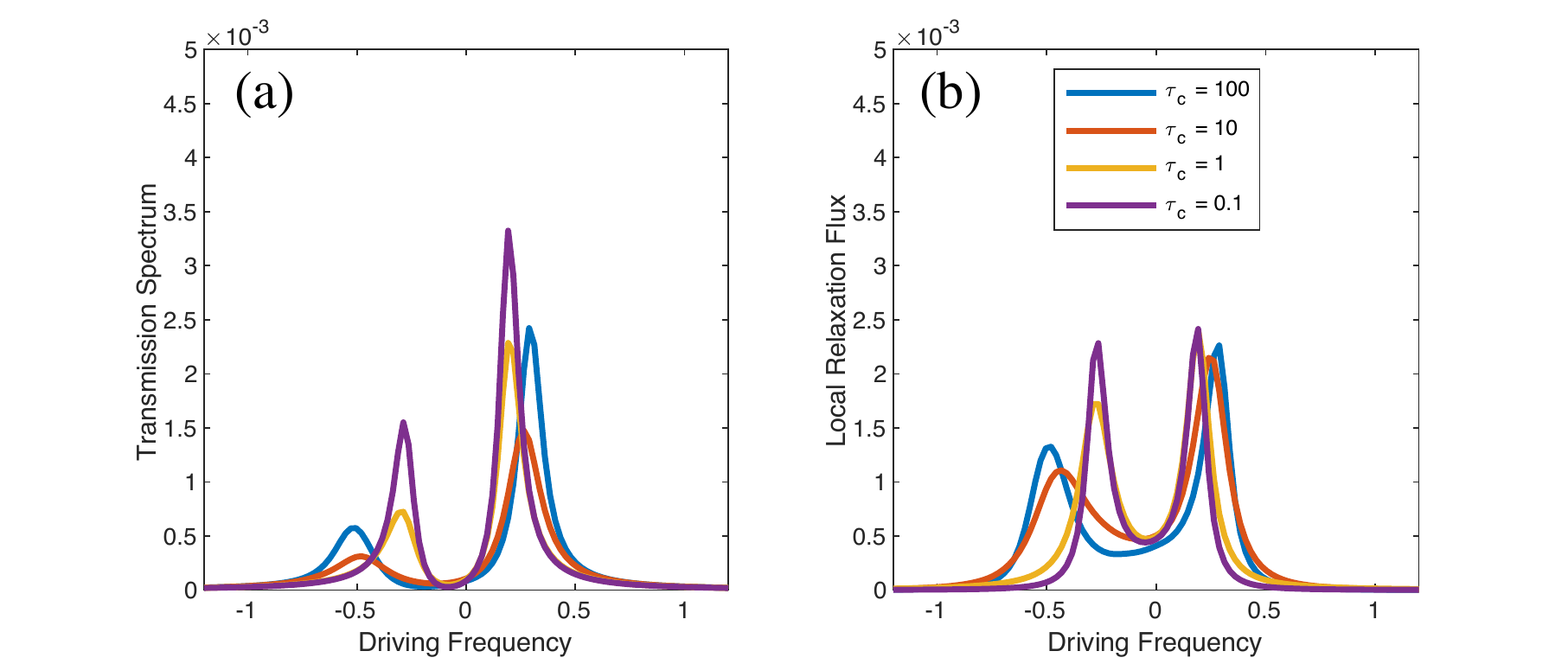}
\caption{Steady state (a) transmission and (b) local relaxation flux for an ensemble of emitters inside the cavity with intermolecular couplings and dynamic disorder. Both figures are for the indicated correlation times but a fixed disorder strength ($\Delta = 0.4\pi$). Again, the long correlation time limit (blue lines) is equivalent to static disorder results (yellow lines in Fig. \ref{fig: incavdipsd_transvschem}). Both the Rabi splitting and the linewidth decreases as the correlation time decreases similar to Fig. \ref{fig: incavddtranschemangle}.  The center of the two polariton peaks shift to lower energy by the intermolecular coupling. Same as in Fig. \ref{fig: incavdipsd_transvschem}, the difference in height between the upper and lower polariton peaks is more significant in the transmission spectrum than in the local relaxation signal.}
\label{fig: incavdipdd_transvschem}
\end{figure}
 
With these structures in mind, 
in Fig~\ref{fig: outcavsdddangulardip}, we plot the absorption spectra for different orientational disorder strengths. When there is no orientational disorder, because there is finite negative intermolecular coupling, the peaks are red shifted, as the brightest eigenstate corresponds to the smallest eigenvalue. As we increase the disorder strength, the total absorption decreases. For the dynamic orientational disorder case, similar to Fig. \ref{fig: outcavsdddangular}, as the correlation time decreases, the absorption linewidth decreases (corresponding to motional narrowing). The scale of the red shift is also reduced as we increase modulation speed.

In Fig. \ref{fig: incavdipsd_transvschem}, we plot the transmission and local relaxation flux for an ensemble of emitters placed inside the cavity. 
First, one might guess from what was shown in Fig. \ref{fig: densityofstatesfordifferentprobeangles}, the upper polariton peak height is greater than that of the lower polariton; the center of the two polariton peaks has a lower energy than that of the bare cavity mode. 
Second, as the static disorder strength increases, the effective Rabi splitting decreases because the disorder reduces the collectiveness of the emitters. 
Third, as we compare the two channels, because the transmission spectra relies on the cavity mode and the local relaxation relies on the emitter states, the difference in intensity between the upper and lower polaritons for transmission spectrum is greater than that for the local relaxation signals. 
Lastly, the dark modes contribute only to the local relaxation signals, as was also seen in Fig. \ref{fig: incavsdtranschemangular}.

Finally, we show the steady state transmission and local relaxation flux inside a single-mode cavity with intermolecular coupling and dynamic orientational disorder in Fig. \ref{fig: incavdipdd_transvschem}. As the correlation time decreases, the effective Rabi splitting and the linewidth of each polariton peak decreases, just as in Fig. \ref{fig: incavddtranschemangle}. The difference between the upper and lower polariton peak height is less significant in local relaxation flux than the transmission spectrum, similar to what was found in Fig. \ref{fig: incavdipsd_transvschem}. All of the observations are consistent with Section \ref{subsec: orientationaldisorder}.

\section{cavity effect on local relaxation yield}

The models defined by eqs \ref{eq:1}-\ref{eq: nhpart2} are minimal models that allow considerations of three relaxation channels for the radiation absorbed from the driving field: a radiative channel that combines transmission and reflection, nonradiative damping of the cavity mode (heat production in the cavity walls) and reactive relaxation. The latter represents a reaction that occurs following the molecular excitation. It is of interest to examine the effect of the cavity environment on the yield of the latter channel.
 
Outside the cavity, the system undergoes collective superradiance at a rate $N\Gamma_{rad}$ and local relaxation at a rate $\Gamma_{loc}$. The yield of the individual molecular relaxation (that we assume to be a reactive process) is 
\begin{align}
    \text{Yield} = \frac{\Gamma_{loc}}{N\Gamma_{rad} + \Gamma_{loc}}
    \label{eq: nocav_yield}
\end{align}

By comparison, inside the cavity,  the energy will leak through the cavity mode at rate $\Gamma_{cav}$ with the same local relaxation rate as above,  $\Gamma_{loc}$. It is  clear that the local yield is then
\begin{align}
    \text{Yield} = \frac{\Gamma_{loc}}{\Gamma_{cav} + \Gamma_{loc}}
    \label{eq: cav_yield}
\end{align}

\begin{figure}
\includegraphics[width=15cm]{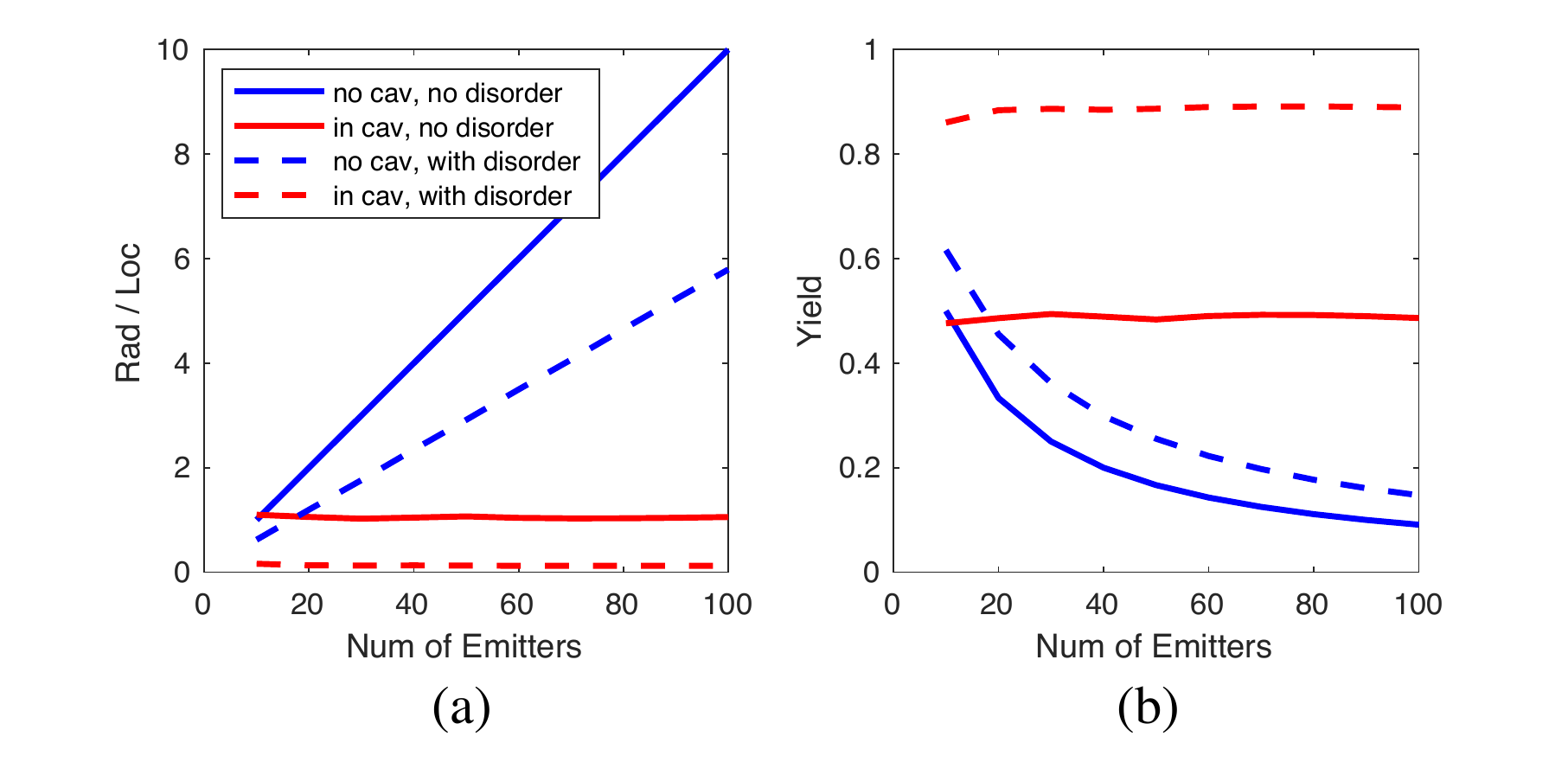}
\caption{(a) The steady state ratio between radiative flux and local relaxation flux outside/inside the cavity with/without energy dynamic disorder. (b) The steady state local relaxation yield outside/inside cavity with/without energy dynamic disorder. We keep all parameters constant and vary only the number of emitters of the system. We assign the cavity leakage rate to be the same as the local relaxation rate for the case of the system inside the cavity. For the case without disorder, we choose the superradiant rate for $N = 10$ to be the same as the local relaxation rate for the case outside the cavity. Hence, the local yield at $N = 10$ is 0.5 both inside and outside the cavity. For outside the cavity, as more emitters are collectively emitting, the ratio between the radiative and local relaxation rate increases linearly as $N$, and thus the local yield decreases. For inside the cavity, we assume the cavity leakage rate remains approximately constant. Hence, the local relaxation yield does not depend on $N$.\cite{bisht2018collective}
From this data, we conclude that adding a single mode cavity can certainly change the importance of different energy dissipation channels.
Now, for the case with moderately fast energetic dynamic disorder ($\sigma = 0.2$, $\tau_{c} = 1$), one clear pattern is that for both outside/inside the cavity, the radiative channels are suppressed by the disorder, and thus, local relaxation yields increase.
Moreover, for the case outside the cavity (blue dashed line), the slope decreases from $0.1$ (no disorder) to $0.06$ (moderate dynamic disorder), which shows that there are 60\% emitters behaving collectively. 
In applying this conclusion to  a realistic experiment, note that $N$ should be the effective number of emitters that participate in the superradiant rate/Rabi splitting, which characterize the collectiveness under the influence of disorder.} 
\label{fig: chemyieldoutincavity}
\end{figure}

Hence, without disorder, we can expect that the reactive relaxation yields will be different outside and inside the cavity, both in terms of absolute value and $N$ dependence. 
This difference is shown in Fig. \ref{fig: chemyieldoutincavity}, 
Note, for simplicity, we assume $\Gamma_{cav} = 10\Gamma_{rad}$.
In agreement with eqs \ref{eq: nocav_yield} and \ref{eq: cav_yield}, the local relaxation yield depends on $N$ only when the emitters are outside the cavity.

Consider now the effect of disorder. As discussed above and shown in Figs. 
\ref{fig: outsidecavitysddd} and \ref{fig: incavddtranschem}, at reasonably high temperatures, nuclear motion and dephasing processes lead to static and dynamic disorder, which results in a reduction of steady state radiative fluxes. As shown in Fig. \ref{fig: chemyieldoutincavity}, we can clearly see that moderate dynamic disorder leads to a reduction of radiative fluxes, and thus a greater local relaxation yield both inside and outside the cavity. 
More importantly, the slope of the ratio between radiative and local relaxation flux (blue-dashed line in the subfigure (a)) is reduced in the presence of disorder outside the cavity. In principle, this slope corresponds to the effective number of emitters that contribute to the collective emission.  For the parameters in Fig. \ref{fig: chemyieldoutincavity}, if we focus on comparing the blue solid and dashed lines, we find a slope of $0.06$ for moderate disorder (blue, dashed line), vs $0.1$ for no disorder case (blue solid line), which implies that 60\% molecules are emitting collectively.  That being said, for very strong dephasing and disorder, we expect that this slope should approach zero and $N$ in eq \ref{eq: nocav_yield} should approach unity; after all, with strong enough dephasing, we expect to find independent spontaneous emission for emitters outside the cavity -- in contrast to the superradiant effect in eq \ref{eq: nocav_yield}. While this analysis demonstrate a possible cavity effect on the yield of chemical reaction, a note of caution should be added. This analysis was based on the assumption that $\Gamma_{loc}$ is identical following excitation of the bright mode outside the cavity as in the case where a polariton is excited inside the cavity. This is not necessarily the case because the initially prepared state differs in energy by an amount determined by the Rabi splitting. A model where $\Gamma_{loc}$ is sensitive to this difference was recently analyzed in ref \citenum{cui2022collective}

\section{Conclusion}
In conclusion, we investigate the influence of energetic disorder, orientational disorder and intermolecular couplings on the absorption spectra for an ensemble of two-level emitters outside and inside a single-mode cavity. As far as the influence of energetic and orientational disorder are concerned, we recover inhomogeneous broadening for static disorder and motional narrowing for dynamic disorder. For the case inside the cavity, there are two inherently different observables: transmission spectrum and local relaxation flux. The transmission spectrum is dictated primarily by properties of the cavity mode and the local relaxation flux depends mainly on the property of the individual emitters. These two observables are complementary to each other and together provide a comprehensive picture of the system when we introduce different types of disorder. Introducing intermolecular coupling is essentially equivalent to adding finite detuning between the bare cavity mode and the emitters eigenstates. As we compare the two observables, we confirm that dark modes do not contribute to transmission signals which requires the state to be bright. However, the dark modes do contribute to the local relaxation signals.

Lastly, we investigate the differences between excited state local relaxation yields  outside versus inside the cavity. It is clear that the presence of a cavity mode will modify the yield in two ways: (1) changing the radiative energy dissipation rate (from collective spontaneous emission to cavity leakage), and (2) changing inherently how collectiveness translates into physical observables (from superradiance to Rabi splitting)\cite{bisht2018collective}.

Looking forward, we note that, in this work, we have assumed that the rate of independent spontaneous emission towards other polarization direction is very slow and can be completely ignored. 
In truth, however, the influence of other polarization directions cannot really be modeled with a single-mode cavity. To better isolate possible cavity effects in chemistry,  one of our next steps will be  to incorporate higher dimensions and more cavity modes inside a realistic microcavity (likely through a Maxwell-Bloch calculation).

\appendix
\section{Generation of Orenstein-Uhlenbeck process\label{apdx: ouprocess}}
In our computational procedure, for generating energy modulations, we employ an Orenstein-Uhlenbeck process. To generate a sequence of values at time $t_{n} = t_{0}+ndt$, we sample $\{\Omega_{m}(t_{n})\}$ in the following process:\cite{Rybicki1994notes}

\begin{enumerate}
    \item At $t_0$, pick $\{\Omega_{m}(t_{0})\}$ according to Gaussian distribution
    $$    P[\Omega_{m}(t_{0})] = \frac{1}{\sqrt{2\pi\sigma^2}}\exp\Bigl(-\frac{\Omega_{m}(t_{0})^2}{2\sigma^2}\Bigr)$$
    \item Calculate $r_{n}$:
    $$r_{n} = \exp\bigl(-(t_{n+1} - t_{n})/\tau_{c}\bigr) = \exp\bigl(-dt/\tau_{c}\bigr)$$
    \item Pick $\{\Omega_{m}(t_{n+1})\}$ according to conditional probability 
    $$P[\Omega_{m}(t_{n + 1})|\Omega_{m}(t_{n})]=\frac{1}{\sqrt{2\pi\sigma^2(1-r_{n}^2)}}\exp\Bigl(-\frac{(\Omega_{m}(t_{n+1})-r_{n}\Omega_{m}(t_{n}))^2}{2\sigma^2(1-r_{n}^2)}\Bigr)$$
    \item Go back to 2
\end{enumerate}
\section{Voigt theory\label{apdx: voigt_theory}}
\begin{figure}[!ht]
\includegraphics[width=15cm]{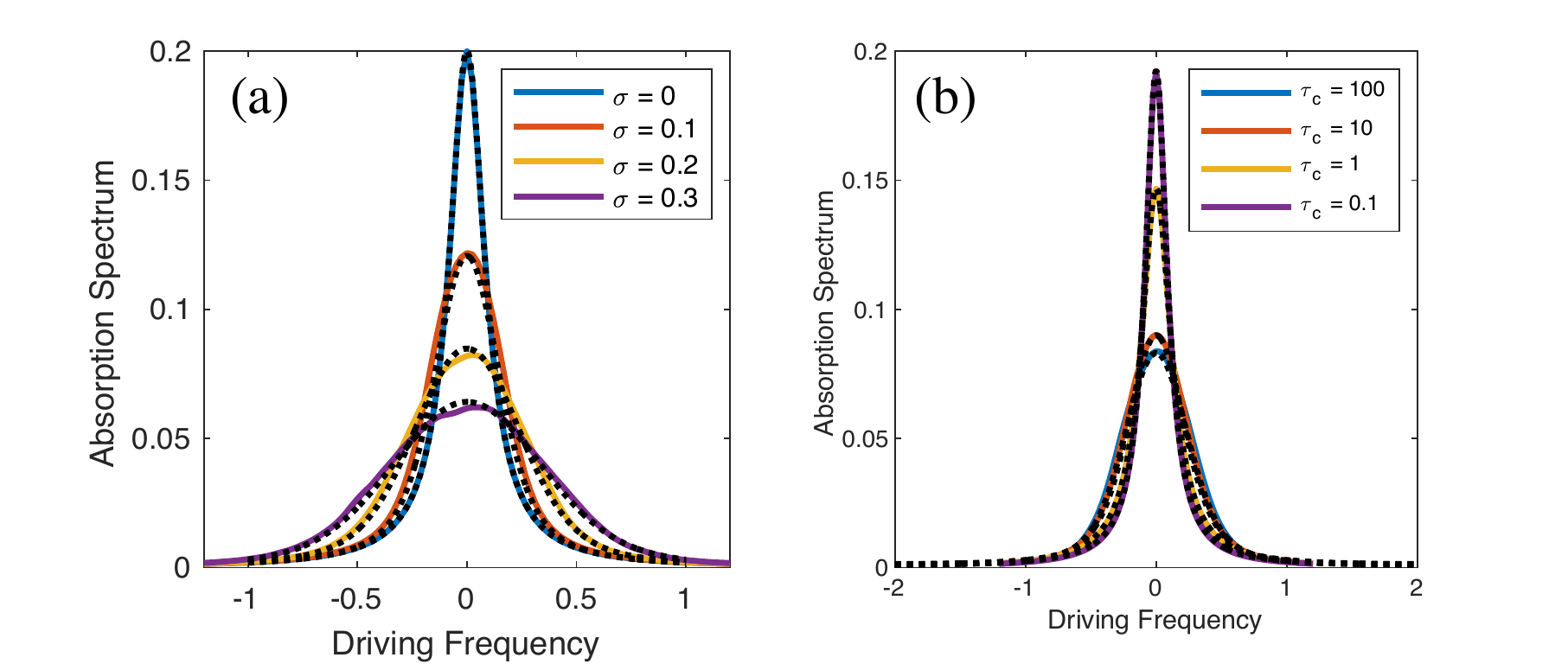}
\caption{Steady state absorption spectrum for a molecular ensemble with (a) different static disorder strengths and with (b) different correlation time $\tau_{c}$ but the same disorder strength $\sigma = 0.2$, identical to Fig. \ref{fig: outsidecavitysddd}. The black dots are predicted by eq \ref{eq: sdvoigt} and \ref{eq: ddvoigt}. The linewidths of numerical results match perfectly with the prediction.}
\label{fig: outsidecavitysdddwithvoigt}
\end{figure}

In this appendix, we demonstrate that our numerical results from section \ref{subsubsec: resnocavsddd} for the case of a collection of molecules outside the cavity with static/dynamic energetic disorder (Fig. \ref{fig: outsidecavitysddd}) satisfy Voigt theory. According to eq \ref{eq: ddvoigt} and eq \ref{eq: sdvoigt}, we can analytically predict the Voigt
linewidth for different disorder strengths $\sigma$ and correlation time $\tau_{c}$. However, we cannot determine the absolute peak height from eq \ref{eq: ddvoigt} and eq \ref{eq: sdvoigt}.  Nevertheless, if we simply fit the lineshape with the numerical height, then Fig. \ref{fig: outsidecavitysdddwithvoigt} demonstrates that the anatlyic linewidths predicted by eq \ref{eq: sdvoigt} and eq \ref{eq: ddvoigt} (lack dots) match perfectly with the numerical linewidths in Fig. \ref{fig: outsidecavitysddd}.

\section{Derivation of the Effective Rabi Splitting under Static Energetic Disorder\label{apdx: sdfitting}}
In this appendix, we use second-order perturbation theory to derive eq \ref{eq: rabisplittingvssigma}.
The total Hamiltonian of the disordered system is
\begin{align}
    \hat{H}_{sub} =& H_{0} + \Lambda
    \label{eq: subblockham}
    \\
    H_{0} =& E_{v}\ketbra{v}{v} + \sum_{m=1}^{N}E_{m}\ketbra{m}{m} + V_{mv}(\ketbra{m}{v} + \ketbra{v}{m})
    \\
    \Lambda =& \sum_{m=1}^{N}\delta{E}_{m}\ketbra{m}{m}
    \label{eq: subblockham3}
\end{align}
Here $N$ is the effective number of emitters in the cavity. The modulation $\delta E_{m}$ is treated as perturbation and has zero mean and $\sigma^2$ variance. We assume no detuning and thus, the Rabi splitting with no disorder is
\begin{equation}
    \Omega_{R}(\sigma=0)=2\sqrt{N}V_{mv}
\end{equation}
The polariton states have energy $\pm\Omega_{R}(\sigma=0) / 2$ and the corresponding eigenvectors are
\begin{align}
    \ket{UP/LP} = (1/\sqrt{2}, \pm1/\sqrt{2N}, ..., \pm1/\sqrt{2N})
    \label{eq: lpupdef}
\end{align}
All remaining eigenstates are dark states and have energy $E_{dark}=0$.
Second-order perturbation theory yields the energies of the polariton states
\begin{align}
    E_{LP} = E_{LP}^{0}+\bra{LP}\Lambda\ket{LP}+\sum_{k\neq LP}\frac{|\bra{k}\Lambda\ket{LP}|^2}{E_{LP}^{0} - E_{k}^{0}}
\end{align}
The first-order term is the mean of the static disorder.
\begin{align}
    \bra{LP}\Lambda\ket{LP} = \sum_{m = 1}^{N}\delta{E}_{m}/2N = 0
\end{align}
The second term is
\begin{align}
    \sum_{k\neq LP}\frac{|\bra{k}\Lambda\ket{LP}|^2}{E_{LP}^{0} - E_{k}^{0}} =& -\frac{\sum_{k }|\bra{k}\Lambda\ket{LP}|^2-|\bra{LP}\Lambda\ket{LP}|^2 - |\bra{UP}\Lambda\ket{LP}|^2/2}{\Omega_{R}(\sigma_{0})/2}
    \\
    \sum_{k}|\bra{k}\Lambda\ket{LP}|^2 =& \bra{LP}\Lambda^2\ket{LP}=\sum_{m=1}^{N}\delta E_{m}^2/2N = \sigma^2/2
    \\
    |\bra{UP}\Lambda\ket{LP}|^2/2 = & -\sum_{m=1}^{N}\delta E_{m} / 4N = 0
\end{align}
Hence, the perturbed energy is
\begin{align}
    E_{LP} = E_{LP}^{0}-\frac{\sigma^2}{\Omega_{R}(\sigma=0)}
\end{align}
Last, because the detuning is zero, the spectrum is completely symmetrical. The effective Rabi splitting is simply twice the absolute value of the lower polariton energy, which yields eq \ref{eq: rabisplittingvssigma}.
\section{Derivation of the Effective Rabi Splitting under Static and Dynamic Angular Disorder\label{apdx: saddadtheory}}
\begin{figure}[!ht]
\includegraphics[width=15cm]{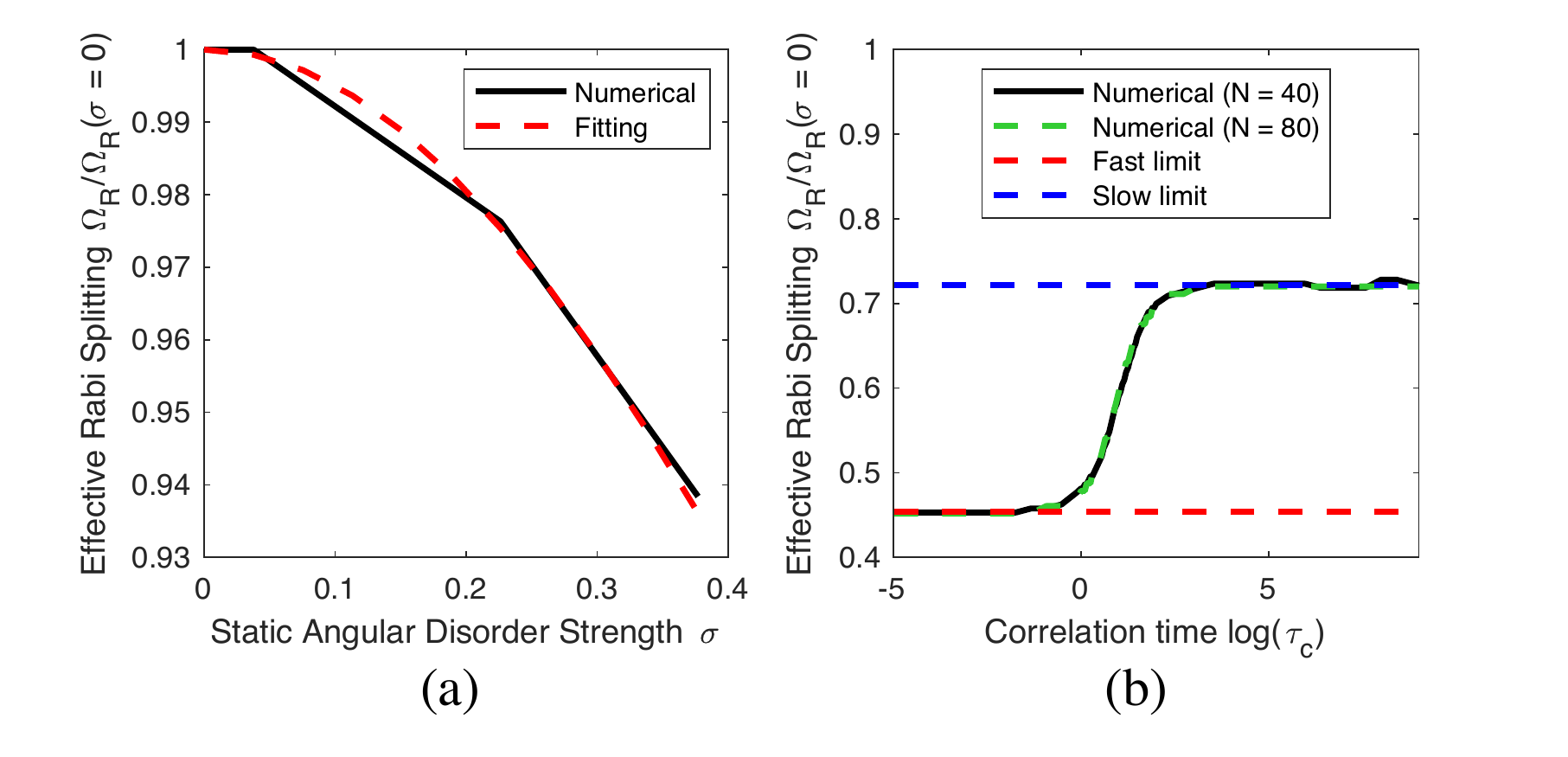}
\caption{Effective Rabi splitting vs static angular disorder strength and dynamic angular disorder correlation time. As shown in figure (a), static angular disorder leads to contraction of Rabi splitting. This contraction is unlike the case with static energetic disorder, in which the effective Rabi splitting increases as the disorder strength increases. The formula eq \ref{eq: saders} obtained by cumulant expansion captures the contraction of effective Rabi splitting induced by static angular disorder ($\tau_{c} \rightarrow \infty$). For the case with dynamic angular disorder ($\sigma = 0.4\pi$), as shown by the Red dashed line in figure (b), 
we can only predict the fast modulation limit $\tau_{c}=0$ as in eq \ref{eq: rabisplittingfastmodulation}. For the intermediate correlation time, the effective Rabi splitting is not fully determined by the first and second moment. As shown by the green dashed figure, the effective Rabi splitting is also independent of $N$ in the system.
\label{fig: sadfitting}}
\end{figure}

In this appendix, we use a cumulant expansion to calculate the effective Rabi splitting that is relevant with disorder as a function of the disorder strength $\sigma$ and/or correlation time $\tau_{c}$.
At this point, if we ignore the pumping and non-hermitian decay from eqs \ref{eq: hamincav} - \ref{eq: nhpart2}, the Hamiltonian of the disordered system is
\begin{align}
    \hat{H} =& H_{0} + \Lambda_{\theta}
    \\
    H_{0} =& E_{v}\ketbra{v}{v} + \sum_{m=1}^{N}E_{m}\ketbra{m}{m}
    \\
    \Lambda_{\theta} =& \sum_{m=1}^{N}( V_{mv}(\theta_{m})\ketbra{m}{v} + V_{vm}(\theta_{m})\ketbra{v}{m}
    \label{eq: }
\end{align}
As shown in eq \ref{eq: staticangulardisordercos}, dynamic angular disorder enters as a modulation of $\theta$ in the coupling $V_{mv}(\theta_{m}) = V\cos(\theta_{m})$ between each individual emitter and the bare cavity mode. It is straightforward to obtain the instantaneous eigenvalues of $H$ because the characteristic polynomial is
\begin{align}
    \lambda^{N-1}(\lambda^{2} - \sum_{m = 1}^{N}V^{2}\cos^{2}(\theta_{m})) = 0
\end{align}
Hence, the effective Rabi splitting is $\Omega_{R}/\Omega_{R}(\sigma = 0) = \sqrt{\langle\cos^{2}\theta_{m}\rangle}$, where ($\theta_{m}(t)$) satisfies $\langle\theta_{m}(t)\rangle = 0$ and $\langle\theta_{m}(t_{j})\theta_{n}(t_{k})\rangle = \delta_{mn}\sigma^2\exp(-|t_{j}-t_{k}|/\tau_{c})$. By cumulant expansion
\begin{align}
    \langle \cos{\theta_{m}}\rangle =& \text{Re}(\exp(i\langle \theta_{m}\rangle - \frac{1}{2}\langle\delta \theta_{m}^{2}\rangle)) = \exp(-\sigma^2/2)
    \label{eq: rabisplittingslowmod}
\end{align}
Hence, in the static limit ($\tau_{c} = \infty$), 
\begin{align}
    \langle\cos^{2} \theta_{m} \rangle = \Bigl\langle\frac{\cos2\theta_{m}+1}{2}\Bigr\rangle= \frac{(e^{-2\sigma^2} + 1)}{2}
\end{align}
The effective Rabi splitting is
\begin{align}
    \Omega_{R}(\sigma, \tau_{c} = \infty) / \Omega_{R}(\sigma = 0)=\sqrt{(e^{-2\sigma^2} + 1)/2}
    \label{eq: saders}
\end{align}
This analytical formula is verified in Fig. \ref{fig: sadfitting}. The numerical results are extracted from Fig. \ref{fig: incavsdtranschemangular}.

In the fast modulation limit ($\tau_{c}=0$), the external field sees an averaged upper/lower polariton state (as defined in eq \ref{eq: lpupdef} in Appendix \ref{apdx: sdfitting}), which leads to average Rabi splitting
\begin{align}
    \Omega_{R}(\sigma, \tau_{c} = 0) / \Omega_{R}(\sigma = 0)=\langle\cos \theta_{m} \rangle = e^{-\sigma^2/2}
    \label{eq: rabisplittingfastmodulation}
\end{align}

Unfortunately, we have not yet been able to derive an analytic form for the effective Rabi splitting for finite $\tau_{c}$. For a general $\tau_c$, the effective Rabi splitting is not determined exclusively by the instantaneous eigenvalues (which allowed us to calculate the slow and fast limits (eqs. \ref{eq: saders} and \ref{eq: rabisplittingfastmodulation}) above).  That being said, in Fig. \ref{fig: sadfitting}, we plot the effective Rabi splitting for two different values of N (the number of emitters). Note that the two lines are on top of one another. Thus, even though we do not have a predictive theory for the Rabi splitting in the intermediate $\tau_c$ regime, we can at least be confident that the Rabi splitting does always scale as $\sqrt{N}$ (which would seem to agree with the results in ref. \citenum{perez2022effective}).

\section{Symbols, energies/rates/frequencies and values that appear in this paper}
In this Appendix, we will list all parameters in the paper. 
\begin{table}
  \begin{threeparttable}[b]
   \caption[]{Symbols, energies/rates/frequencies, and values that appear in this paper.}
   \centering
   \begin{tabular}{cP{8cm}c}
     \midrule 
     Symbol&    Energy/Rate/Frequency & Value
    \\\hline
    $E_{v}$   & Cavity mode energy & 0\tnote{*}
    \\
    $E_{0}$ & Total ground state energy&-1\tnote{*}
    \\
    $E_{m}$ & molecular singly excited state energy& 0 (for no disorder system)
    \\
    $\hat{V}_{mv}$ & coupling between cavity mode $\ket{v}$ and molecular excited state $\ket{m}$& $0.5/\sqrt{N}$\tnote{$\dagger$}
    \\
    $\hat{V}_{m0}$ & coupling between ground state $\ket{0}$ and molecular excited state $\ket{m}$& 0.0005
    \\
    $\omega$ & Driving frequency & [-1.2, 1.2]
    \\
    $\Gamma_{rad}$ & Single molecule radiative decay rate & 0.05/N
    \\
    $\Gamma_{loc}$ & Local relaxation rate & 0.05
    \\
    $\Gamma_{cav} (\Gamma_{cav}^{R}+\Gamma_{cav}^{NR}) $ & Cavity leakage rate (including radiative and nonradiative) & 0.05
    \\
    \midrule
     \end{tabular}
     \begin{tablenotes}
       \item [$\dagger$] We choose to normalize the Rabi splitting for no disorder system and thus all parameters are in unit of $\Omega_{R}(\sigma = 0)$.
       \item [*] The choice of energy difference between ground and cavity mode energy/molecular singly excited state energy does not affect the results because the calculation is under rotating wave approximation. We also set the energy of the singly excited state energy/cavity mode energy for no disorder system to be $0$.
     \end{tablenotes}
  \end{threeparttable}
\end{table}

\bibliography{apssamp}

\begin{thebibliography}{62}%
\makeatletter
\providecommand \@ifxundefined [1]{%
 \@ifx{#1\undefined}
}%
\providecommand \@ifnum [1]{%
 \ifnum #1\expandafter \@firstoftwo
 \else \expandafter \@secondoftwo
 \fi
}%
\providecommand \@ifx [1]{%
 \ifx #1\expandafter \@firstoftwo
 \else \expandafter \@secondoftwo
 \fi
}%
\providecommand \natexlab [1]{#1}%
\providecommand \enquote  [1]{``#1''}%
\providecommand \bibnamefont  [1]{#1}%
\providecommand \bibfnamefont [1]{#1}%
\providecommand \citenamefont [1]{#1}%
\providecommand \href@noop [0]{\@secondoftwo}%
\providecommand \href [0]{\begingroup \@sanitize@url \@href}%
\providecommand \@href[1]{\@@startlink{#1}\@@href}%
\providecommand \@@href[1]{\endgroup#1\@@endlink}%
\providecommand \@sanitize@url [0]{\catcode `\\12\catcode `\$12\catcode
  `\&12\catcode `\#12\catcode `\^12\catcode `\_12\catcode `\%12\relax}%
\providecommand \@@startlink[1]{}%
\providecommand \@@endlink[0]{}%
\providecommand \url  [0]{\begingroup\@sanitize@url \@url }%
\providecommand \@url [1]{\endgroup\@href {#1}{\urlprefix }}%
\providecommand \urlprefix  [0]{URL }%
\providecommand \Eprint [0]{\href }%
\providecommand \doibase [0]{https://doi.org/}%
\providecommand \selectlanguage [0]{\@gobble}%
\providecommand \bibinfo  [0]{\@secondoftwo}%
\providecommand \bibfield  [0]{\@secondoftwo}%
\providecommand \translation [1]{[#1]}%
\providecommand \BibitemOpen [0]{}%
\providecommand \bibitemStop [0]{}%
\providecommand \bibitemNoStop [0]{.\EOS\space}%
\providecommand \EOS [0]{\spacefactor3000\relax}%
\providecommand \BibitemShut  [1]{\csname bibitem#1\endcsname}%
\let\auto@bib@innerbib\@empty
\bibitem [{\citenamefont {Dicke}(1954)}]{dicke1954coherence}%
  \BibitemOpen
  \bibfield  {author} {\bibinfo {author} {\bibfnamefont {R.~H.}\ \bibnamefont
  {Dicke}},\ }\bibfield  {title} {\enquote {\bibinfo {title} {Coherence in
  spontaneous radiation processes},}\ }\href@noop {} {\bibfield  {journal}
  {\bibinfo  {journal} {Physical review}\ }\textbf {\bibinfo {volume} {93}},\
  \bibinfo {pages} {99} (\bibinfo {year} {1954})}\BibitemShut {NoStop}%
\bibitem [{\citenamefont {Spano}\ and\ \citenamefont
  {Mukamel}(1989)}]{spano_superradiance_1989}%
  \BibitemOpen
  \bibfield  {author} {\bibinfo {author} {\bibfnamefont {F.~C.}\ \bibnamefont
  {Spano}}\ and\ \bibinfo {author} {\bibfnamefont {S.}~\bibnamefont
  {Mukamel}},\ }\bibfield  {title} {\enquote {\bibinfo {title} {Superradiance
  in molecular aggregates},}\ }\href {https://doi.org/10.1063/1.457174}
  {\bibfield  {journal} {\bibinfo  {journal} {The Journal of Chemical Physics}\
  }\textbf {\bibinfo {volume} {91}},\ \bibinfo {pages} {683--700} (\bibinfo
  {year} {1989})}\BibitemShut {NoStop}%
\bibitem [{\citenamefont {Lim}\ \emph {et~al.}(2004)\citenamefont {Lim},
  \citenamefont {Bjorklund}, \citenamefont {Spano},\ and\ \citenamefont
  {Bardeen}}]{lim_exciton_2004}%
  \BibitemOpen
  \bibfield  {author} {\bibinfo {author} {\bibfnamefont {S.-H.}\ \bibnamefont
  {Lim}}, \bibinfo {author} {\bibfnamefont {T.~G.}\ \bibnamefont {Bjorklund}},
  \bibinfo {author} {\bibfnamefont {F.~C.}\ \bibnamefont {Spano}},\ and\
  \bibinfo {author} {\bibfnamefont {C.~J.}\ \bibnamefont {Bardeen}},\
  }\bibfield  {title} {\enquote {\bibinfo {title} {Exciton {Delocalization} and
  {Superradiance} in {Tetracene} {Thin} {Films} and {Nanoaggregates}},}\
  }\href@noop {} {\bibfield  {journal} {\bibinfo  {journal} {Physical Review
  Letters}\ }\textbf {\bibinfo {volume} {92}},\ \bibinfo {pages} {107402}
  (\bibinfo {year} {2004})}\BibitemShut {NoStop}%
\bibitem [{\citenamefont {Bonifacio}\ and\ \citenamefont
  {Lugiato}(1975)}]{bonifacio1975cooperative}%
  \BibitemOpen
  \bibfield  {author} {\bibinfo {author} {\bibfnamefont {R.}~\bibnamefont
  {Bonifacio}}\ and\ \bibinfo {author} {\bibfnamefont {L.}~\bibnamefont
  {Lugiato}},\ }\bibfield  {title} {\enquote {\bibinfo {title} {Cooperative
  radiation processes in two-level systems: Superfluorescence},}\ }\href@noop
  {} {\bibfield  {journal} {\bibinfo  {journal} {Physical Review A}\ }\textbf
  {\bibinfo {volume} {11}},\ \bibinfo {pages} {1507} (\bibinfo {year}
  {1975})}\BibitemShut {NoStop}%
\bibitem [{\citenamefont {Rain{\`o}}\ \emph {et~al.}(2018)\citenamefont
  {Rain{\`o}}, \citenamefont {Becker}, \citenamefont {Bodnarchuk},
  \citenamefont {Mahrt}, \citenamefont {Kovalenko},\ and\ \citenamefont
  {St{\"o}ferle}}]{raino2018superfluorescence}%
  \BibitemOpen
  \bibfield  {author} {\bibinfo {author} {\bibfnamefont {G.}~\bibnamefont
  {Rain{\`o}}}, \bibinfo {author} {\bibfnamefont {M.~A.}\ \bibnamefont
  {Becker}}, \bibinfo {author} {\bibfnamefont {M.~I.}\ \bibnamefont
  {Bodnarchuk}}, \bibinfo {author} {\bibfnamefont {R.~F.}\ \bibnamefont
  {Mahrt}}, \bibinfo {author} {\bibfnamefont {M.~V.}\ \bibnamefont
  {Kovalenko}},\ and\ \bibinfo {author} {\bibfnamefont {T.}~\bibnamefont
  {St{\"o}ferle}},\ }\bibfield  {title} {\enquote {\bibinfo {title}
  {Superfluorescence from lead halide perovskite quantum dot superlattices},}\
  }\href@noop {} {\bibfield  {journal} {\bibinfo  {journal} {Nature}\ }\textbf
  {\bibinfo {volume} {563}},\ \bibinfo {pages} {671--675} (\bibinfo {year}
  {2018})}\BibitemShut {NoStop}%
\bibitem [{\citenamefont {Raimond}\ \emph {et~al.}(1982)\citenamefont
  {Raimond}, \citenamefont {Goy}, \citenamefont {Gross}, \citenamefont
  {Fabre},\ and\ \citenamefont {Haroche}}]{raimond1982collective}%
  \BibitemOpen
  \bibfield  {author} {\bibinfo {author} {\bibfnamefont {J.}~\bibnamefont
  {Raimond}}, \bibinfo {author} {\bibfnamefont {P.}~\bibnamefont {Goy}},
  \bibinfo {author} {\bibfnamefont {M.}~\bibnamefont {Gross}}, \bibinfo
  {author} {\bibfnamefont {C.}~\bibnamefont {Fabre}},\ and\ \bibinfo {author}
  {\bibfnamefont {S.}~\bibnamefont {Haroche}},\ }\bibfield  {title} {\enquote
  {\bibinfo {title} {Collective absorption of blackbody radiation by rydberg
  atoms in a cavity: an experiment on bose statistics and brownian motion},}\
  }\href@noop {} {\bibfield  {journal} {\bibinfo  {journal} {Physical Review
  Letters}\ }\textbf {\bibinfo {volume} {49}},\ \bibinfo {pages} {117}
  (\bibinfo {year} {1982})}\BibitemShut {NoStop}%
\bibitem [{\citenamefont {Weisbuch}\ \emph {et~al.}(1992)\citenamefont
  {Weisbuch}, \citenamefont {Nishioka}, \citenamefont {Ishikawa},\ and\
  \citenamefont {Arakawa}}]{weisbuch1992observation}%
  \BibitemOpen
  \bibfield  {author} {\bibinfo {author} {\bibfnamefont {C.}~\bibnamefont
  {Weisbuch}}, \bibinfo {author} {\bibfnamefont {M.}~\bibnamefont {Nishioka}},
  \bibinfo {author} {\bibfnamefont {A.}~\bibnamefont {Ishikawa}},\ and\
  \bibinfo {author} {\bibfnamefont {Y.}~\bibnamefont {Arakawa}},\ }\bibfield
  {title} {\enquote {\bibinfo {title} {Observation of the coupled
  exciton-photon mode splitting in a semiconductor quantum microcavity},}\
  }\href@noop {} {\bibfield  {journal} {\bibinfo  {journal} {Physical Review
  Letters}\ }\textbf {\bibinfo {volume} {69}},\ \bibinfo {pages} {3314}
  (\bibinfo {year} {1992})}\BibitemShut {NoStop}%
\bibitem [{\citenamefont {Heinzen}\ \emph {et~al.}(1987)\citenamefont
  {Heinzen}, \citenamefont {Childs}, \citenamefont {Thomas},\ and\
  \citenamefont {Feld}}]{heinzen1987enhanced}%
  \BibitemOpen
  \bibfield  {author} {\bibinfo {author} {\bibfnamefont {D.}~\bibnamefont
  {Heinzen}}, \bibinfo {author} {\bibfnamefont {J.}~\bibnamefont {Childs}},
  \bibinfo {author} {\bibfnamefont {J.}~\bibnamefont {Thomas}},\ and\ \bibinfo
  {author} {\bibfnamefont {M.}~\bibnamefont {Feld}},\ }\bibfield  {title}
  {\enquote {\bibinfo {title} {Enhanced and inhibited visible spontaneous
  emission by atoms in a confocal resonator},}\ }\href@noop {} {\bibfield
  {journal} {\bibinfo  {journal} {Physical review letters}\ }\textbf {\bibinfo
  {volume} {58}},\ \bibinfo {pages} {1320} (\bibinfo {year}
  {1987})}\BibitemShut {NoStop}%
\bibitem [{\citenamefont {Jhe}\ \emph {et~al.}(1987)\citenamefont {Jhe},
  \citenamefont {Anderson}, \citenamefont {Hinds}, \citenamefont {Meschede},
  \citenamefont {Moi},\ and\ \citenamefont {Haroche}}]{jhe1987suppression}%
  \BibitemOpen
  \bibfield  {author} {\bibinfo {author} {\bibfnamefont {W.}~\bibnamefont
  {Jhe}}, \bibinfo {author} {\bibfnamefont {A.}~\bibnamefont {Anderson}},
  \bibinfo {author} {\bibfnamefont {E.~A.}\ \bibnamefont {Hinds}}, \bibinfo
  {author} {\bibfnamefont {D.}~\bibnamefont {Meschede}}, \bibinfo {author}
  {\bibfnamefont {L.}~\bibnamefont {Moi}},\ and\ \bibinfo {author}
  {\bibfnamefont {S.}~\bibnamefont {Haroche}},\ }\bibfield  {title} {\enquote
  {\bibinfo {title} {Suppression of spontaneous decay at optical frequencies:
  Test of vacuum-field anisotropy in confined space},}\ }\href@noop {}
  {\bibfield  {journal} {\bibinfo  {journal} {Physical review letters}\
  }\textbf {\bibinfo {volume} {58}},\ \bibinfo {pages} {666} (\bibinfo {year}
  {1987})}\BibitemShut {NoStop}%
\bibitem [{\citenamefont {De~Martini}\ \emph {et~al.}(1987)\citenamefont
  {De~Martini}, \citenamefont {Innocenti}, \citenamefont {Jacobovitz},\ and\
  \citenamefont {Mataloni}}]{de1987anomalous}%
  \BibitemOpen
  \bibfield  {author} {\bibinfo {author} {\bibfnamefont {F.}~\bibnamefont
  {De~Martini}}, \bibinfo {author} {\bibfnamefont {G.}~\bibnamefont
  {Innocenti}}, \bibinfo {author} {\bibfnamefont {G.}~\bibnamefont
  {Jacobovitz}},\ and\ \bibinfo {author} {\bibfnamefont {P.}~\bibnamefont
  {Mataloni}},\ }\bibfield  {title} {\enquote {\bibinfo {title} {Anomalous
  spontaneous emission time in a microscopic optical cavity},}\ }\href@noop {}
  {\bibfield  {journal} {\bibinfo  {journal} {Physical review letters}\
  }\textbf {\bibinfo {volume} {59}},\ \bibinfo {pages} {2955} (\bibinfo {year}
  {1987})}\BibitemShut {NoStop}%
\bibitem [{\citenamefont {Bayer}\ \emph {et~al.}(2001)\citenamefont {Bayer},
  \citenamefont {Reinecke}, \citenamefont {Weidner}, \citenamefont {Larionov},
  \citenamefont {McDonald},\ and\ \citenamefont
  {Forchel}}]{bayer2001inhibition}%
  \BibitemOpen
  \bibfield  {author} {\bibinfo {author} {\bibfnamefont {M.}~\bibnamefont
  {Bayer}}, \bibinfo {author} {\bibfnamefont {T.~L.}\ \bibnamefont {Reinecke}},
  \bibinfo {author} {\bibfnamefont {F.}~\bibnamefont {Weidner}}, \bibinfo
  {author} {\bibfnamefont {A.}~\bibnamefont {Larionov}}, \bibinfo {author}
  {\bibfnamefont {A.}~\bibnamefont {McDonald}},\ and\ \bibinfo {author}
  {\bibfnamefont {A.}~\bibnamefont {Forchel}},\ }\bibfield  {title} {\enquote
  {\bibinfo {title} {Inhibition and enhancement of the spontaneous emission of
  quantum dots in structured microresonators},}\ }\href@noop {} {\bibfield
  {journal} {\bibinfo  {journal} {Physical review letters}\ }\textbf {\bibinfo
  {volume} {86}},\ \bibinfo {pages} {3168} (\bibinfo {year}
  {2001})}\BibitemShut {NoStop}%
\bibitem [{\citenamefont {Houdr{\'e}}\ \emph {et~al.}(1994)\citenamefont
  {Houdr{\'e}}, \citenamefont {Stanley}, \citenamefont {Oesterle},
  \citenamefont {Ilegems},\ and\ \citenamefont {Weisbuch}}]{houdre1994room}%
  \BibitemOpen
  \bibfield  {author} {\bibinfo {author} {\bibfnamefont {R.}~\bibnamefont
  {Houdr{\'e}}}, \bibinfo {author} {\bibfnamefont {R.}~\bibnamefont {Stanley}},
  \bibinfo {author} {\bibfnamefont {U.}~\bibnamefont {Oesterle}}, \bibinfo
  {author} {\bibfnamefont {M.}~\bibnamefont {Ilegems}},\ and\ \bibinfo {author}
  {\bibfnamefont {C.}~\bibnamefont {Weisbuch}},\ }\bibfield  {title} {\enquote
  {\bibinfo {title} {Room-temperature cavity polaritons in a semiconductor
  microcavity},}\ }\href@noop {} {\bibfield  {journal} {\bibinfo  {journal}
  {Physical Review B}\ }\textbf {\bibinfo {volume} {49}},\ \bibinfo {pages}
  {16761} (\bibinfo {year} {1994})}\BibitemShut {NoStop}%
\bibitem [{\citenamefont {G{\'e}rard}\ \emph {et~al.}(1998)\citenamefont
  {G{\'e}rard}, \citenamefont {Sermage}, \citenamefont {Gayral}, \citenamefont
  {Legrand}, \citenamefont {Costard},\ and\ \citenamefont
  {Thierry-Mieg}}]{gerard1998enhanced}%
  \BibitemOpen
  \bibfield  {author} {\bibinfo {author} {\bibfnamefont {J.-M.}\ \bibnamefont
  {G{\'e}rard}}, \bibinfo {author} {\bibfnamefont {B.}~\bibnamefont {Sermage}},
  \bibinfo {author} {\bibfnamefont {B.}~\bibnamefont {Gayral}}, \bibinfo
  {author} {\bibfnamefont {B.}~\bibnamefont {Legrand}}, \bibinfo {author}
  {\bibfnamefont {E.}~\bibnamefont {Costard}},\ and\ \bibinfo {author}
  {\bibfnamefont {V.}~\bibnamefont {Thierry-Mieg}},\ }\bibfield  {title}
  {\enquote {\bibinfo {title} {Enhanced spontaneous emission by quantum boxes
  in a monolithic optical microcavity},}\ }\href@noop {} {\bibfield  {journal}
  {\bibinfo  {journal} {Physical review letters}\ }\textbf {\bibinfo {volume}
  {81}},\ \bibinfo {pages} {1110} (\bibinfo {year} {1998})}\BibitemShut
  {NoStop}%
\bibitem [{\citenamefont {Yoshie}\ \emph {et~al.}(2004)\citenamefont {Yoshie},
  \citenamefont {Scherer}, \citenamefont {Hendrickson}, \citenamefont
  {Khitrova}, \citenamefont {Gibbs}, \citenamefont {Rupper}, \citenamefont
  {Ell}, \citenamefont {Shchekin},\ and\ \citenamefont
  {Deppe}}]{yoshie2004vacuum}%
  \BibitemOpen
  \bibfield  {author} {\bibinfo {author} {\bibfnamefont {T.}~\bibnamefont
  {Yoshie}}, \bibinfo {author} {\bibfnamefont {A.}~\bibnamefont {Scherer}},
  \bibinfo {author} {\bibfnamefont {J.}~\bibnamefont {Hendrickson}}, \bibinfo
  {author} {\bibfnamefont {G.}~\bibnamefont {Khitrova}}, \bibinfo {author}
  {\bibfnamefont {H.}~\bibnamefont {Gibbs}}, \bibinfo {author} {\bibfnamefont
  {G.}~\bibnamefont {Rupper}}, \bibinfo {author} {\bibfnamefont
  {C.}~\bibnamefont {Ell}}, \bibinfo {author} {\bibfnamefont {O.}~\bibnamefont
  {Shchekin}},\ and\ \bibinfo {author} {\bibfnamefont {D.}~\bibnamefont
  {Deppe}},\ }\bibfield  {title} {\enquote {\bibinfo {title} {Vacuum rabi
  splitting with a single quantum dot in a photonic crystal nanocavity},}\
  }\href@noop {} {\bibfield  {journal} {\bibinfo  {journal} {Nature}\ }\textbf
  {\bibinfo {volume} {432}},\ \bibinfo {pages} {200--203} (\bibinfo {year}
  {2004})}\BibitemShut {NoStop}%
\bibitem [{\citenamefont {Masiello}\ and\ \citenamefont
  {Schatz}(2008)}]{masiello2008many}%
  \BibitemOpen
  \bibfield  {author} {\bibinfo {author} {\bibfnamefont {D.~J.}\ \bibnamefont
  {Masiello}}\ and\ \bibinfo {author} {\bibfnamefont {G.~C.}\ \bibnamefont
  {Schatz}},\ }\bibfield  {title} {\enquote {\bibinfo {title} {Many-body theory
  of surface-enhanced raman scattering},}\ }\href@noop {} {\bibfield  {journal}
  {\bibinfo  {journal} {Physical Review A}\ }\textbf {\bibinfo {volume} {78}},\
  \bibinfo {pages} {042505} (\bibinfo {year} {2008})}\BibitemShut {NoStop}%
\bibitem [{\citenamefont {Spano}(2010)}]{spano_spectral_2010}%
  \BibitemOpen
  \bibfield  {author} {\bibinfo {author} {\bibfnamefont {F.~C.}\ \bibnamefont
  {Spano}},\ }\bibfield  {title} {\enquote {\bibinfo {title} {The {Spectral}
  {Signatures} of {Frenkel} {Polarons} in {H}- and {J}-{Aggregates}},}\
  }\href@noop {} {\bibfield  {journal} {\bibinfo  {journal} {Accounts of
  Chemical Research}\ }\textbf {\bibinfo {volume} {43}},\ \bibinfo {pages}
  {429--439} (\bibinfo {year} {2010})}\BibitemShut {NoStop}%
\bibitem [{\citenamefont {Mirsaleh-Kohan}\ \emph {et~al.}(2012)\citenamefont
  {Mirsaleh-Kohan}, \citenamefont {Iberi}, \citenamefont {Simmons~Jr},
  \citenamefont {Bigelow}, \citenamefont {Vaschillo}, \citenamefont {Rowland},
  \citenamefont {Best}, \citenamefont {Pennycook}, \citenamefont {Masiello},
  \citenamefont {Guiton} \emph {et~al.}}]{mirsaleh2012single}%
  \BibitemOpen
  \bibfield  {author} {\bibinfo {author} {\bibfnamefont {N.}~\bibnamefont
  {Mirsaleh-Kohan}}, \bibinfo {author} {\bibfnamefont {V.}~\bibnamefont
  {Iberi}}, \bibinfo {author} {\bibfnamefont {P.~D.}\ \bibnamefont
  {Simmons~Jr}}, \bibinfo {author} {\bibfnamefont {N.~W.}\ \bibnamefont
  {Bigelow}}, \bibinfo {author} {\bibfnamefont {A.}~\bibnamefont {Vaschillo}},
  \bibinfo {author} {\bibfnamefont {M.~M.}\ \bibnamefont {Rowland}}, \bibinfo
  {author} {\bibfnamefont {M.~D.}\ \bibnamefont {Best}}, \bibinfo {author}
  {\bibfnamefont {S.~J.}\ \bibnamefont {Pennycook}}, \bibinfo {author}
  {\bibfnamefont {D.~J.}\ \bibnamefont {Masiello}}, \bibinfo {author}
  {\bibfnamefont {B.~S.}\ \bibnamefont {Guiton}}, \emph {et~al.},\ }\bibfield
  {title} {\enquote {\bibinfo {title} {Single-molecule surface-enhanced raman
  scattering: can stem/eels image electromagnetic hot spots?}}\ }\href@noop {}
  {\bibfield  {journal} {\bibinfo  {journal} {The Journal of Physical Chemistry
  Letters}\ }\textbf {\bibinfo {volume} {3}},\ \bibinfo {pages} {2303--2309}
  (\bibinfo {year} {2012})}\BibitemShut {NoStop}%
\bibitem [{\citenamefont {Spano}(2015)}]{spano_optical_2015}%
  \BibitemOpen
  \bibfield  {author} {\bibinfo {author} {\bibfnamefont {F.~C.}\ \bibnamefont
  {Spano}},\ }\bibfield  {title} {\enquote {\bibinfo {title} {Optical
  microcavities enhance the exciton coherence length and eliminate vibronic
  coupling in {J}-aggregates},}\ }\href@noop {} {\bibfield  {journal} {\bibinfo
   {journal} {The Journal of Chemical Physics}\ }\textbf {\bibinfo {volume}
  {142}},\ \bibinfo {pages} {184707} (\bibinfo {year} {2015})}\BibitemShut
  {NoStop}%
\bibitem [{\citenamefont {Herrera}\ and\ \citenamefont
  {Spano}(2016)}]{herrera_cavity_controlled_2016}%
  \BibitemOpen
  \bibfield  {author} {\bibinfo {author} {\bibfnamefont {F.}~\bibnamefont
  {Herrera}}\ and\ \bibinfo {author} {\bibfnamefont {F.~C.}\ \bibnamefont
  {Spano}},\ }\bibfield  {title} {\enquote {\bibinfo {title}
  {Cavity-{Controlled} {Chemistry} in {Molecular} {Ensembles}},}\ }\href@noop
  {} {\bibfield  {journal} {\bibinfo  {journal} {Physical Review Letters}\
  }\textbf {\bibinfo {volume} {116}} (\bibinfo {year} {2016})}\BibitemShut
  {NoStop}%
\bibitem [{\citenamefont {Herrera}\ and\ \citenamefont
  {Spano}(2017)}]{herrera_absorption_2017}%
  \BibitemOpen
  \bibfield  {author} {\bibinfo {author} {\bibfnamefont {F.}~\bibnamefont
  {Herrera}}\ and\ \bibinfo {author} {\bibfnamefont {F.~C.}\ \bibnamefont
  {Spano}},\ }\bibfield  {title} {\enquote {\bibinfo {title} {Absorption and
  photoluminescence in organic cavity {QED}},}\ }\href@noop {} {\bibfield
  {journal} {\bibinfo  {journal} {Physical Review A}\ }\textbf {\bibinfo
  {volume} {95}} (\bibinfo {year} {2017})}\BibitemShut {NoStop}%
\bibitem [{\citenamefont {Abid}\ \emph {et~al.}(2017)\citenamefont {Abid},
  \citenamefont {Chen}, \citenamefont {Yuan}, \citenamefont {Bohloul},
  \citenamefont {Najmaei}, \citenamefont {Avendano}, \citenamefont
  {P{\'e}chou}, \citenamefont {Mlayah},\ and\ \citenamefont
  {Lou}}]{abid2017temperature}%
  \BibitemOpen
  \bibfield  {author} {\bibinfo {author} {\bibfnamefont {I.}~\bibnamefont
  {Abid}}, \bibinfo {author} {\bibfnamefont {W.}~\bibnamefont {Chen}}, \bibinfo
  {author} {\bibfnamefont {J.}~\bibnamefont {Yuan}}, \bibinfo {author}
  {\bibfnamefont {A.}~\bibnamefont {Bohloul}}, \bibinfo {author} {\bibfnamefont
  {S.}~\bibnamefont {Najmaei}}, \bibinfo {author} {\bibfnamefont
  {C.}~\bibnamefont {Avendano}}, \bibinfo {author} {\bibfnamefont
  {R.}~\bibnamefont {P{\'e}chou}}, \bibinfo {author} {\bibfnamefont
  {A.}~\bibnamefont {Mlayah}},\ and\ \bibinfo {author} {\bibfnamefont
  {J.}~\bibnamefont {Lou}},\ }\bibfield  {title} {\enquote {\bibinfo {title}
  {Temperature-dependent plasmon--exciton interactions in hybrid au/mose2
  nanostructures},}\ }\href@noop {} {\bibfield  {journal} {\bibinfo  {journal}
  {ACS photonics}\ }\textbf {\bibinfo {volume} {4}},\ \bibinfo {pages}
  {1653--1660} (\bibinfo {year} {2017})}\BibitemShut {NoStop}%
\bibitem [{\citenamefont {Bisht}\ \emph {et~al.}(2018)\citenamefont {Bisht},
  \citenamefont {Cuadra}, \citenamefont {Wersall}, \citenamefont {Canales},
  \citenamefont {Antosiewicz},\ and\ \citenamefont
  {Shegai}}]{bisht2018collective}%
  \BibitemOpen
  \bibfield  {author} {\bibinfo {author} {\bibfnamefont {A.}~\bibnamefont
  {Bisht}}, \bibinfo {author} {\bibfnamefont {J.}~\bibnamefont {Cuadra}},
  \bibinfo {author} {\bibfnamefont {M.}~\bibnamefont {Wersall}}, \bibinfo
  {author} {\bibfnamefont {A.}~\bibnamefont {Canales}}, \bibinfo {author}
  {\bibfnamefont {T.~J.}\ \bibnamefont {Antosiewicz}},\ and\ \bibinfo {author}
  {\bibfnamefont {T.}~\bibnamefont {Shegai}},\ }\bibfield  {title} {\enquote
  {\bibinfo {title} {Collective strong light-matter coupling in hierarchical
  microcavity-plasmon-exciton systems},}\ }\href@noop {} {\bibfield  {journal}
  {\bibinfo  {journal} {Nano letters}\ }\textbf {\bibinfo {volume} {19}},\
  \bibinfo {pages} {189--196} (\bibinfo {year} {2018})}\BibitemShut {NoStop}%
\bibitem [{\citenamefont {Hertzog}\ \emph {et~al.}(2019)\citenamefont
  {Hertzog}, \citenamefont {Wang}, \citenamefont {Mony},\ and\ \citenamefont
  {B{\"o}rjesson}}]{hertzog2019strong}%
  \BibitemOpen
  \bibfield  {author} {\bibinfo {author} {\bibfnamefont {M.}~\bibnamefont
  {Hertzog}}, \bibinfo {author} {\bibfnamefont {M.}~\bibnamefont {Wang}},
  \bibinfo {author} {\bibfnamefont {J.}~\bibnamefont {Mony}},\ and\ \bibinfo
  {author} {\bibfnamefont {K.}~\bibnamefont {B{\"o}rjesson}},\ }\bibfield
  {title} {\enquote {\bibinfo {title} {Strong light--matter interactions: a new
  direction within chemistry},}\ }\href@noop {} {\bibfield  {journal} {\bibinfo
   {journal} {Chemical Society Reviews}\ }\textbf {\bibinfo {volume} {48}},\
  \bibinfo {pages} {937--961} (\bibinfo {year} {2019})}\BibitemShut {NoStop}%
\bibitem [{\citenamefont {Sidler}\ \emph {et~al.}(2020)\citenamefont {Sidler},
  \citenamefont {Schäfer}, \citenamefont {Ruggenthaler},\ and\ \citenamefont
  {Rubio}}]{sidler2020polaritonic}%
  \BibitemOpen
  \bibfield  {author} {\bibinfo {author} {\bibfnamefont {D.}~\bibnamefont
  {Sidler}}, \bibinfo {author} {\bibfnamefont {C.}~\bibnamefont {Schäfer}},
  \bibinfo {author} {\bibfnamefont {M.}~\bibnamefont {Ruggenthaler}},\ and\
  \bibinfo {author} {\bibfnamefont {A.}~\bibnamefont {Rubio}},\ }\bibfield
  {title} {\enquote {\bibinfo {title} {Polaritonic chemistry: Collective strong
  coupling implies strong local modification of chemical properties},}\
  }\href@noop {} {\bibfield  {journal} {\bibinfo  {journal} {The journal of
  physical chemistry letters}\ }\textbf {\bibinfo {volume} {12}},\ \bibinfo
  {pages} {508--516} (\bibinfo {year} {2020})}\BibitemShut {NoStop}%
\bibitem [{\citenamefont {Smith}, \citenamefont {Bhattacharya},\ and\
  \citenamefont {Masiello}(2021)}]{smith2021exact}%
  \BibitemOpen
  \bibfield  {author} {\bibinfo {author} {\bibfnamefont {K.~C.}\ \bibnamefont
  {Smith}}, \bibinfo {author} {\bibfnamefont {A.}~\bibnamefont
  {Bhattacharya}},\ and\ \bibinfo {author} {\bibfnamefont {D.~J.}\ \bibnamefont
  {Masiello}},\ }\bibfield  {title} {\enquote {\bibinfo {title} {Exact k-body
  representation of the jaynes-cummings interaction in the dressed basis:
  Insight into many-body phenomena with light},}\ }\href@noop {} {\bibfield
  {journal} {\bibinfo  {journal} {Physical Review A}\ }\textbf {\bibinfo
  {volume} {104}},\ \bibinfo {pages} {013707} (\bibinfo {year}
  {2021})}\BibitemShut {NoStop}%
\bibitem [{\citenamefont {Li}\ \emph {et~al.}(2022)\citenamefont {Li},
  \citenamefont {Cui}, \citenamefont {Subotnik},\ and\ \citenamefont
  {Nitzan}}]{li2022molecular}%
  \BibitemOpen
  \bibfield  {author} {\bibinfo {author} {\bibfnamefont {T.~E.}\ \bibnamefont
  {Li}}, \bibinfo {author} {\bibfnamefont {B.}~\bibnamefont {Cui}}, \bibinfo
  {author} {\bibfnamefont {J.~E.}\ \bibnamefont {Subotnik}},\ and\ \bibinfo
  {author} {\bibfnamefont {A.}~\bibnamefont {Nitzan}},\ }\bibfield  {title}
  {\enquote {\bibinfo {title} {Molecular polaritonics: chemical dynamics under
  strong light--matter coupling},}\ }\href@noop {} {\bibfield  {journal}
  {\bibinfo  {journal} {Annual review of physical chemistry}\ }\textbf
  {\bibinfo {volume} {73}},\ \bibinfo {pages} {43--71} (\bibinfo {year}
  {2022})}\BibitemShut {NoStop}%
\bibitem [{\citenamefont {Renaud}, \citenamefont {Whitley},\ and\ \citenamefont
  {Stroud~Jr}(1977)}]{renaud1977nonstationary}%
  \BibitemOpen
  \bibfield  {author} {\bibinfo {author} {\bibfnamefont {B.}~\bibnamefont
  {Renaud}}, \bibinfo {author} {\bibfnamefont {R.}~\bibnamefont {Whitley}},\
  and\ \bibinfo {author} {\bibfnamefont {C.}~\bibnamefont {Stroud~Jr}},\
  }\bibfield  {title} {\enquote {\bibinfo {title} {Nonstationary two-level
  resonance fluorescence},}\ }\href@noop {} {\bibfield  {journal} {\bibinfo
  {journal} {Journal of Physics B: Atomic and Molecular Physics (1968-1987)}\
  }\textbf {\bibinfo {volume} {10}},\ \bibinfo {pages} {19} (\bibinfo {year}
  {1977})}\BibitemShut {NoStop}%
\bibitem [{\citenamefont {Gross}\ and\ \citenamefont
  {Haroche}(1982)}]{gross1982superradiance}%
  \BibitemOpen
  \bibfield  {author} {\bibinfo {author} {\bibfnamefont {M.}~\bibnamefont
  {Gross}}\ and\ \bibinfo {author} {\bibfnamefont {S.}~\bibnamefont
  {Haroche}},\ }\bibfield  {title} {\enquote {\bibinfo {title} {Superradiance:
  An essay on the theory of collective spontaneous emission},}\ }\href@noop {}
  {\bibfield  {journal} {\bibinfo  {journal} {Physics reports}\ }\textbf
  {\bibinfo {volume} {93}},\ \bibinfo {pages} {301--396} (\bibinfo {year}
  {1982})}\BibitemShut {NoStop}%
\bibitem [{\citenamefont {Celardo}\ \emph {et~al.}(2013)\citenamefont
  {Celardo}, \citenamefont {Biella}, \citenamefont {Kaplan},\ and\
  \citenamefont {Borgonovi}}]{celardo2013interplay}%
  \BibitemOpen
  \bibfield  {author} {\bibinfo {author} {\bibfnamefont {G.}~\bibnamefont
  {Celardo}}, \bibinfo {author} {\bibfnamefont {A.}~\bibnamefont {Biella}},
  \bibinfo {author} {\bibfnamefont {L.}~\bibnamefont {Kaplan}},\ and\ \bibinfo
  {author} {\bibfnamefont {F.}~\bibnamefont {Borgonovi}},\ }\bibfield  {title}
  {\enquote {\bibinfo {title} {Interplay of superradiance and disorder in the
  anderson model},}\ }\href@noop {} {\bibfield  {journal} {\bibinfo  {journal}
  {Fortschritte der Physik}\ }\textbf {\bibinfo {volume} {61}},\ \bibinfo
  {pages} {250--260} (\bibinfo {year} {2013})}\BibitemShut {NoStop}%
\bibitem [{\citenamefont {Biella}\ \emph
  {et~al.}(2013{\natexlab{a}})\citenamefont {Biella}, \citenamefont
  {Borgonovi}, \citenamefont {Kaiser},\ and\ \citenamefont
  {Celardo}}]{biella_subradiant_2013}%
  \BibitemOpen
  \bibfield  {author} {\bibinfo {author} {\bibfnamefont {A.}~\bibnamefont
  {Biella}}, \bibinfo {author} {\bibfnamefont {F.}~\bibnamefont {Borgonovi}},
  \bibinfo {author} {\bibfnamefont {R.}~\bibnamefont {Kaiser}},\ and\ \bibinfo
  {author} {\bibfnamefont {G.~L.}\ \bibnamefont {Celardo}},\ }\bibfield
  {title} {\enquote {\bibinfo {title} {Subradiant hybrid states in the open
  {3D} {Anderson}-{Dicke} model},}\ }\href@noop {} {\bibfield  {journal}
  {\bibinfo  {journal} {EPL (Europhysics Letters)}\ }\textbf {\bibinfo {volume}
  {103}},\ \bibinfo {pages} {57009} (\bibinfo {year}
  {2013}{\natexlab{a}})}\BibitemShut {NoStop}%
\bibitem [{\citenamefont {Biella}\ \emph
  {et~al.}(2013{\natexlab{b}})\citenamefont {Biella}, \citenamefont
  {Borgonovi}, \citenamefont {Kaiser},\ and\ \citenamefont
  {Celardo}}]{biella2013subradiant}%
  \BibitemOpen
  \bibfield  {author} {\bibinfo {author} {\bibfnamefont {A.}~\bibnamefont
  {Biella}}, \bibinfo {author} {\bibfnamefont {F.}~\bibnamefont {Borgonovi}},
  \bibinfo {author} {\bibfnamefont {R.}~\bibnamefont {Kaiser}},\ and\ \bibinfo
  {author} {\bibfnamefont {G.}~\bibnamefont {Celardo}},\ }\bibfield  {title}
  {\enquote {\bibinfo {title} {Subradiant hybrid states in the open 3d
  anderson-dicke model},}\ }\href@noop {} {\bibfield  {journal} {\bibinfo
  {journal} {EPL (Europhysics Letters)}\ }\textbf {\bibinfo {volume} {103}},\
  \bibinfo {pages} {57009} (\bibinfo {year} {2013}{\natexlab{b}})}\BibitemShut
  {NoStop}%
\bibitem [{\citenamefont {Delga}\ \emph {et~al.}(2014)\citenamefont {Delga},
  \citenamefont {Feist}, \citenamefont {Bravo-Abad},\ and\ \citenamefont
  {Garcia-Vidal}}]{delga2014quantum}%
  \BibitemOpen
  \bibfield  {author} {\bibinfo {author} {\bibfnamefont {A.}~\bibnamefont
  {Delga}}, \bibinfo {author} {\bibfnamefont {J.}~\bibnamefont {Feist}},
  \bibinfo {author} {\bibfnamefont {J.}~\bibnamefont {Bravo-Abad}},\ and\
  \bibinfo {author} {\bibfnamefont {F.}~\bibnamefont {Garcia-Vidal}},\
  }\bibfield  {title} {\enquote {\bibinfo {title} {Quantum emitters near a
  metal nanoparticle: strong coupling and quenching},}\ }\href@noop {}
  {\bibfield  {journal} {\bibinfo  {journal} {Physical review letters}\
  }\textbf {\bibinfo {volume} {112}},\ \bibinfo {pages} {253601} (\bibinfo
  {year} {2014})}\BibitemShut {NoStop}%
\bibitem [{\citenamefont {Celardo}, \citenamefont {Giusteri},\ and\
  \citenamefont {Borgonovi}(2014)}]{celardo2014cooperative}%
  \BibitemOpen
  \bibfield  {author} {\bibinfo {author} {\bibfnamefont {G.~L.}\ \bibnamefont
  {Celardo}}, \bibinfo {author} {\bibfnamefont {G.~G.}\ \bibnamefont
  {Giusteri}},\ and\ \bibinfo {author} {\bibfnamefont {F.}~\bibnamefont
  {Borgonovi}},\ }\bibfield  {title} {\enquote {\bibinfo {title} {Cooperative
  robustness to static disorder: Superradiance and localization in a nanoscale
  ring to model light-harvesting systems found in nature},}\ }\href@noop {}
  {\bibfield  {journal} {\bibinfo  {journal} {Physical Review B}\ }\textbf
  {\bibinfo {volume} {90}},\ \bibinfo {pages} {075113} (\bibinfo {year}
  {2014})}\BibitemShut {NoStop}%
\bibitem [{\citenamefont {Pustovit}, \citenamefont {Urbas},\ and\ \citenamefont
  {Shahbazyan}(2014)}]{pustovit2014energy}%
  \BibitemOpen
  \bibfield  {author} {\bibinfo {author} {\bibfnamefont {V.~N.}\ \bibnamefont
  {Pustovit}}, \bibinfo {author} {\bibfnamefont {A.~M.}\ \bibnamefont
  {Urbas}},\ and\ \bibinfo {author} {\bibfnamefont {T.~V.}\ \bibnamefont
  {Shahbazyan}},\ }\bibfield  {title} {\enquote {\bibinfo {title} {Energy
  transfer in plasmonic systems},}\ }\href@noop {} {\bibfield  {journal}
  {\bibinfo  {journal} {Journal of Optics}\ }\textbf {\bibinfo {volume} {16}},\
  \bibinfo {pages} {114015} (\bibinfo {year} {2014})}\BibitemShut {NoStop}%
\bibitem [{\citenamefont {Goban}\ \emph {et~al.}(2015)\citenamefont {Goban},
  \citenamefont {Hung}, \citenamefont {Hood}, \citenamefont {Yu}, \citenamefont
  {Muniz}, \citenamefont {Painter},\ and\ \citenamefont
  {Kimble}}]{goban2015superradiance}%
  \BibitemOpen
  \bibfield  {author} {\bibinfo {author} {\bibfnamefont {A.}~\bibnamefont
  {Goban}}, \bibinfo {author} {\bibfnamefont {C.-L.}\ \bibnamefont {Hung}},
  \bibinfo {author} {\bibfnamefont {J.}~\bibnamefont {Hood}}, \bibinfo {author}
  {\bibfnamefont {S.-P.}\ \bibnamefont {Yu}}, \bibinfo {author} {\bibfnamefont
  {J.}~\bibnamefont {Muniz}}, \bibinfo {author} {\bibfnamefont
  {O.}~\bibnamefont {Painter}},\ and\ \bibinfo {author} {\bibfnamefont
  {H.}~\bibnamefont {Kimble}},\ }\bibfield  {title} {\enquote {\bibinfo {title}
  {Superradiance for atoms trapped along a photonic crystal waveguide},}\
  }\href@noop {} {\bibfield  {journal} {\bibinfo  {journal} {Physical review
  letters}\ }\textbf {\bibinfo {volume} {115}},\ \bibinfo {pages} {063601}
  (\bibinfo {year} {2015})}\BibitemShut {NoStop}%
\bibitem [{\citenamefont {Norcia}\ \emph {et~al.}(2016)\citenamefont {Norcia},
  \citenamefont {Winchester}, \citenamefont {Cline},\ and\ \citenamefont
  {Thompson}}]{norcia2016superradiance}%
  \BibitemOpen
  \bibfield  {author} {\bibinfo {author} {\bibfnamefont {M.~A.}\ \bibnamefont
  {Norcia}}, \bibinfo {author} {\bibfnamefont {M.~N.}\ \bibnamefont
  {Winchester}}, \bibinfo {author} {\bibfnamefont {J.~R.}\ \bibnamefont
  {Cline}},\ and\ \bibinfo {author} {\bibfnamefont {J.~K.}\ \bibnamefont
  {Thompson}},\ }\bibfield  {title} {\enquote {\bibinfo {title} {Superradiance
  on the millihertz linewidth strontium clock transition},}\ }\href@noop {}
  {\bibfield  {journal} {\bibinfo  {journal} {Science advances}\ }\textbf
  {\bibinfo {volume} {2}},\ \bibinfo {pages} {e1601231} (\bibinfo {year}
  {2016})}\BibitemShut {NoStop}%
\bibitem [{\citenamefont {Asenjo-Garcia}\ \emph {et~al.}(2017)\citenamefont
  {Asenjo-Garcia}, \citenamefont {Moreno-Cardoner}, \citenamefont {Albrecht},
  \citenamefont {Kimble},\ and\ \citenamefont {Chang}}]{asenjo2017exponential}%
  \BibitemOpen
  \bibfield  {author} {\bibinfo {author} {\bibfnamefont {A.}~\bibnamefont
  {Asenjo-Garcia}}, \bibinfo {author} {\bibfnamefont {M.}~\bibnamefont
  {Moreno-Cardoner}}, \bibinfo {author} {\bibfnamefont {A.}~\bibnamefont
  {Albrecht}}, \bibinfo {author} {\bibfnamefont {H.}~\bibnamefont {Kimble}},\
  and\ \bibinfo {author} {\bibfnamefont {D.~E.}\ \bibnamefont {Chang}},\
  }\bibfield  {title} {\enquote {\bibinfo {title} {Exponential improvement in
  photon storage fidelities using subradiance and “selective radiance” in
  atomic arrays},}\ }\href@noop {} {\bibfield  {journal} {\bibinfo  {journal}
  {Physical Review X}\ }\textbf {\bibinfo {volume} {7}},\ \bibinfo {pages}
  {031024} (\bibinfo {year} {2017})}\BibitemShut {NoStop}%
\bibitem [{\citenamefont {Kirton}\ and\ \citenamefont
  {Keeling}(2017)}]{kirton2017suppressing}%
  \BibitemOpen
  \bibfield  {author} {\bibinfo {author} {\bibfnamefont {P.}~\bibnamefont
  {Kirton}}\ and\ \bibinfo {author} {\bibfnamefont {J.}~\bibnamefont
  {Keeling}},\ }\bibfield  {title} {\enquote {\bibinfo {title} {Suppressing and
  restoring the dicke superradiance transition by dephasing and decay},}\
  }\href@noop {} {\bibfield  {journal} {\bibinfo  {journal} {Physical review
  letters}\ }\textbf {\bibinfo {volume} {118}},\ \bibinfo {pages} {123602}
  (\bibinfo {year} {2017})}\BibitemShut {NoStop}%
\bibitem [{\citenamefont {Shahbazyan}, \citenamefont {Raikh},\ and\
  \citenamefont {Vardeny}(2000)}]{shahbazyan2000mesoscopic}%
  \BibitemOpen
  \bibfield  {author} {\bibinfo {author} {\bibfnamefont {T.}~\bibnamefont
  {Shahbazyan}}, \bibinfo {author} {\bibfnamefont {M.}~\bibnamefont {Raikh}},\
  and\ \bibinfo {author} {\bibfnamefont {Z.}~\bibnamefont {Vardeny}},\
  }\bibfield  {title} {\enquote {\bibinfo {title} {Mesoscopic cooperative
  emission from a disordered system},}\ }\href@noop {} {\bibfield  {journal}
  {\bibinfo  {journal} {Physical Review B}\ }\textbf {\bibinfo {volume} {61}},\
  \bibinfo {pages} {13266} (\bibinfo {year} {2000})}\BibitemShut {NoStop}%
\bibitem [{\citenamefont {Sukharev}\ and\ \citenamefont
  {Nitzan}(2017)}]{sukharev2017optics}%
  \BibitemOpen
  \bibfield  {author} {\bibinfo {author} {\bibfnamefont {M.}~\bibnamefont
  {Sukharev}}\ and\ \bibinfo {author} {\bibfnamefont {A.}~\bibnamefont
  {Nitzan}},\ }\bibfield  {title} {\enquote {\bibinfo {title} {Optics of
  exciton-plasmon nanomaterials},}\ }\href@noop {} {\bibfield  {journal}
  {\bibinfo  {journal} {Journal of Physics: Condensed Matter}\ }\textbf
  {\bibinfo {volume} {29}},\ \bibinfo {pages} {443003} (\bibinfo {year}
  {2017})}\BibitemShut {NoStop}%
\bibitem [{\citenamefont {Fauch{\'e}}\ \emph {et~al.}(2017)\citenamefont
  {Fauch{\'e}}, \citenamefont {Gebhardt}, \citenamefont {Sukharev},\ and\
  \citenamefont {Vall{\'e}e}}]{fauche2017plasmonic}%
  \BibitemOpen
  \bibfield  {author} {\bibinfo {author} {\bibfnamefont {P.}~\bibnamefont
  {Fauch{\'e}}}, \bibinfo {author} {\bibfnamefont {C.}~\bibnamefont
  {Gebhardt}}, \bibinfo {author} {\bibfnamefont {M.}~\bibnamefont {Sukharev}},\
  and\ \bibinfo {author} {\bibfnamefont {R.~A.}\ \bibnamefont {Vall{\'e}e}},\
  }\bibfield  {title} {\enquote {\bibinfo {title} {Plasmonic opals: observation
  of a collective molecular exciton mode beyond the strong coupling},}\
  }\href@noop {} {\bibfield  {journal} {\bibinfo  {journal} {Scientific
  reports}\ }\textbf {\bibinfo {volume} {7}},\ \bibinfo {pages} {1--9}
  (\bibinfo {year} {2017})}\BibitemShut {NoStop}%
\bibitem [{\citenamefont {Ribeiro}\ \emph {et~al.}(2018)\citenamefont
  {Ribeiro}, \citenamefont {Mart{\'\i}nez-Mart{\'\i}nez}, \citenamefont {Du},
  \citenamefont {Campos-Gonzalez-Angulo},\ and\ \citenamefont
  {Yuen-Zhou}}]{ribeiro2018polariton}%
  \BibitemOpen
  \bibfield  {author} {\bibinfo {author} {\bibfnamefont {R.~F.}\ \bibnamefont
  {Ribeiro}}, \bibinfo {author} {\bibfnamefont {L.~A.}\ \bibnamefont
  {Mart{\'\i}nez-Mart{\'\i}nez}}, \bibinfo {author} {\bibfnamefont
  {M.}~\bibnamefont {Du}}, \bibinfo {author} {\bibfnamefont {J.}~\bibnamefont
  {Campos-Gonzalez-Angulo}},\ and\ \bibinfo {author} {\bibfnamefont
  {J.}~\bibnamefont {Yuen-Zhou}},\ }\bibfield  {title} {\enquote {\bibinfo
  {title} {Polariton chemistry: controlling molecular dynamics with optical
  cavities},}\ }\href@noop {} {\bibfield  {journal} {\bibinfo  {journal}
  {Chemical science}\ }\textbf {\bibinfo {volume} {9}},\ \bibinfo {pages}
  {6325--6339} (\bibinfo {year} {2018})}\BibitemShut {NoStop}%
\bibitem [{\citenamefont {Herrera}\ and\ \citenamefont
  {Spano}(2018)}]{herrera_theory_2018}%
  \BibitemOpen
  \bibfield  {author} {\bibinfo {author} {\bibfnamefont {F.}~\bibnamefont
  {Herrera}}\ and\ \bibinfo {author} {\bibfnamefont {F.~C.}\ \bibnamefont
  {Spano}},\ }\bibfield  {title} {\enquote {\bibinfo {title} {Theory of
  {Nanoscale} {Organic} {Cavities}: {The} {Essential} {Role} of
  {Vibration}-{Photon} {Dressed} {States}},}\ }\href
  {https://doi.org/10.1021/acsphotonics.7b00728} {\bibfield  {journal}
  {\bibinfo  {journal} {ACS Photonics}\ }\textbf {\bibinfo {volume} {5}},\
  \bibinfo {pages} {65--79} (\bibinfo {year} {2018})}\BibitemShut {NoStop}%
\bibitem [{\citenamefont {G{\'o}mez-Casta{\~n}o}\ \emph
  {et~al.}(2019)\citenamefont {G{\'o}mez-Casta{\~n}o}, \citenamefont
  {Redondo-Cubero}, \citenamefont {Buisson}, \citenamefont {Pau}, \citenamefont
  {Mihi}, \citenamefont {Ravaine}, \citenamefont {Vall{\'e}e}, \citenamefont
  {Nitzan},\ and\ \citenamefont {Sukharev}}]{gomez2019energy}%
  \BibitemOpen
  \bibfield  {author} {\bibinfo {author} {\bibfnamefont {M.}~\bibnamefont
  {G{\'o}mez-Casta{\~n}o}}, \bibinfo {author} {\bibfnamefont {A.}~\bibnamefont
  {Redondo-Cubero}}, \bibinfo {author} {\bibfnamefont {L.}~\bibnamefont
  {Buisson}}, \bibinfo {author} {\bibfnamefont {J.~L.}\ \bibnamefont {Pau}},
  \bibinfo {author} {\bibfnamefont {A.}~\bibnamefont {Mihi}}, \bibinfo {author}
  {\bibfnamefont {S.}~\bibnamefont {Ravaine}}, \bibinfo {author} {\bibfnamefont
  {R.~A.}\ \bibnamefont {Vall{\'e}e}}, \bibinfo {author} {\bibfnamefont
  {A.}~\bibnamefont {Nitzan}},\ and\ \bibinfo {author} {\bibfnamefont
  {M.}~\bibnamefont {Sukharev}},\ }\bibfield  {title} {\enquote {\bibinfo
  {title} {Energy transfer and interference by collective electromagnetic
  coupling},}\ }\href@noop {} {\bibfield  {journal} {\bibinfo  {journal} {Nano
  letters}\ }\textbf {\bibinfo {volume} {19}},\ \bibinfo {pages} {5790--5795}
  (\bibinfo {year} {2019})}\BibitemShut {NoStop}%
\bibitem [{\citenamefont {Wang}, \citenamefont {Scholes},\ and\ \citenamefont
  {Hsu}(2019)}]{wang_quantum_2019}%
  \BibitemOpen
  \bibfield  {author} {\bibinfo {author} {\bibfnamefont {S.}~\bibnamefont
  {Wang}}, \bibinfo {author} {\bibfnamefont {G.~D.}\ \bibnamefont {Scholes}},\
  and\ \bibinfo {author} {\bibfnamefont {L.-Y.}\ \bibnamefont {Hsu}},\
  }\bibfield  {title} {\enquote {\bibinfo {title} {Quantum dynamics of a
  molecular emitter strongly coupled with surface plasmon polaritons: {A}
  macroscopic quantum electrodynamics approach},}\ }\href@noop {} {\bibfield
  {journal} {\bibinfo  {journal} {The Journal of Chemical Physics}\ }\textbf
  {\bibinfo {volume} {151}},\ \bibinfo {pages} {014105} (\bibinfo {year}
  {2019})}\BibitemShut {NoStop}%
\bibitem [{\citenamefont {Wang}, \citenamefont {Scholes},\ and\ \citenamefont
  {Hsu}(2020{\natexlab{a}})}]{wang2020coherent}%
  \BibitemOpen
  \bibfield  {author} {\bibinfo {author} {\bibfnamefont {S.}~\bibnamefont
  {Wang}}, \bibinfo {author} {\bibfnamefont {G.~D.}\ \bibnamefont {Scholes}},\
  and\ \bibinfo {author} {\bibfnamefont {L.-Y.}\ \bibnamefont {Hsu}},\
  }\bibfield  {title} {\enquote {\bibinfo {title} {Coherent-to-incoherent
  transition of molecular fluorescence controlled by surface plasmon
  polaritons},}\ }\href@noop {} {\bibfield  {journal} {\bibinfo  {journal} {The
  journal of physical chemistry letters}\ }\textbf {\bibinfo {volume} {11}},\
  \bibinfo {pages} {5948--5955} (\bibinfo {year}
  {2020}{\natexlab{a}})}\BibitemShut {NoStop}%
\bibitem [{\citenamefont {Mattiotti}\ \emph {et~al.}(2020)\citenamefont
  {Mattiotti}, \citenamefont {Kuno}, \citenamefont {Borgonovi}, \citenamefont
  {Jank{\'o}},\ and\ \citenamefont {Celardo}}]{mattiotti2020thermal}%
  \BibitemOpen
  \bibfield  {author} {\bibinfo {author} {\bibfnamefont {F.}~\bibnamefont
  {Mattiotti}}, \bibinfo {author} {\bibfnamefont {M.}~\bibnamefont {Kuno}},
  \bibinfo {author} {\bibfnamefont {F.}~\bibnamefont {Borgonovi}}, \bibinfo
  {author} {\bibfnamefont {B.}~\bibnamefont {Jank{\'o}}},\ and\ \bibinfo
  {author} {\bibfnamefont {G.~L.}\ \bibnamefont {Celardo}},\ }\bibfield
  {title} {\enquote {\bibinfo {title} {Thermal decoherence of superradiance in
  lead halide perovskite nanocrystal superlattices},}\ }\href@noop {}
  {\bibfield  {journal} {\bibinfo  {journal} {Nano Letters}\ }\textbf {\bibinfo
  {volume} {20}},\ \bibinfo {pages} {7382--7388} (\bibinfo {year}
  {2020})}\BibitemShut {NoStop}%
\bibitem [{\citenamefont {Wang}\ \emph {et~al.}(2020)\citenamefont {Wang},
  \citenamefont {Lee}, \citenamefont {Chuang}, \citenamefont {Scholes},\ and\
  \citenamefont {Hsu}}]{wang_theory_2020}%
  \BibitemOpen
  \bibfield  {author} {\bibinfo {author} {\bibfnamefont {S.}~\bibnamefont
  {Wang}}, \bibinfo {author} {\bibfnamefont {M.-W.}\ \bibnamefont {Lee}},
  \bibinfo {author} {\bibfnamefont {Y.-T.}\ \bibnamefont {Chuang}}, \bibinfo
  {author} {\bibfnamefont {G.~D.}\ \bibnamefont {Scholes}},\ and\ \bibinfo
  {author} {\bibfnamefont {L.-Y.}\ \bibnamefont {Hsu}},\ }\bibfield  {title}
  {\enquote {\bibinfo {title} {Theory of molecular emission power spectra. {I}.
  {Macroscopic} quantum electrodynamics formalism},}\ }\href
  {https://doi.org/10.1063/5.0027796} {\bibfield  {journal} {\bibinfo
  {journal} {The Journal of Chemical Physics}\ }\textbf {\bibinfo {volume}
  {153}},\ \bibinfo {pages} {184102} (\bibinfo {year} {2020})}\BibitemShut
  {NoStop}%
\bibitem [{\citenamefont {Spano}(2020)}]{spano_excitonphonon_2020}%
  \BibitemOpen
  \bibfield  {author} {\bibinfo {author} {\bibfnamefont {F.~C.}\ \bibnamefont
  {Spano}},\ }\bibfield  {title} {\enquote {\bibinfo {title} {Exciton–phonon
  polaritons in organic microcavities: {Testing} a simple ansatz for treating a
  large number of chromophores},}\ }\href@noop {} {\bibfield  {journal}
  {\bibinfo  {journal} {The Journal of Chemical Physics}\ }\textbf {\bibinfo
  {volume} {152}},\ \bibinfo {pages} {204113} (\bibinfo {year}
  {2020})}\BibitemShut {NoStop}%
\bibitem [{\citenamefont {Wang}, \citenamefont {Scholes},\ and\ \citenamefont
  {Hsu}(2020{\natexlab{b}})}]{wang_coherent--incoherent_2020}%
  \BibitemOpen
  \bibfield  {author} {\bibinfo {author} {\bibfnamefont {S.}~\bibnamefont
  {Wang}}, \bibinfo {author} {\bibfnamefont {G.~D.}\ \bibnamefont {Scholes}},\
  and\ \bibinfo {author} {\bibfnamefont {L.-Y.}\ \bibnamefont {Hsu}},\
  }\bibfield  {title} {\enquote {\bibinfo {title} {Coherent-to-{Incoherent}
  {Transition} of {Molecular} {Fluorescence} {Controlled} by {Surface}
  {Plasmon} {Polaritons}},}\ }\href@noop {} {\bibfield  {journal} {\bibinfo
  {journal} {The Journal of Physical Chemistry Letters}\ }\textbf {\bibinfo
  {volume} {11}},\ \bibinfo {pages} {5948--5955} (\bibinfo {year}
  {2020}{\natexlab{b}})}\BibitemShut {NoStop}%
\bibitem [{\citenamefont {Lee}, \citenamefont {Chuang},\ and\ \citenamefont
  {Hsu}(2021)}]{lee_theory_2021}%
  \BibitemOpen
  \bibfield  {author} {\bibinfo {author} {\bibfnamefont {M.-W.}\ \bibnamefont
  {Lee}}, \bibinfo {author} {\bibfnamefont {Y.-T.}\ \bibnamefont {Chuang}},\
  and\ \bibinfo {author} {\bibfnamefont {L.-Y.}\ \bibnamefont {Hsu}},\
  }\bibfield  {title} {\enquote {\bibinfo {title} {Theory of molecular emission
  power spectra. {II}. {Angle}, frequency, and distance dependence of
  electromagnetic environment factor of a molecular emitter in plasmonic
  environments},}\ }\href {https://doi.org/10.1063/5.0057018} {\bibfield
  {journal} {\bibinfo  {journal} {The Journal of Chemical Physics}\ }\textbf
  {\bibinfo {volume} {155}},\ \bibinfo {pages} {074101} (\bibinfo {year}
  {2021})}\BibitemShut {NoStop}%
\bibitem [{\citenamefont {Wang}, \citenamefont {Chuang},\ and\ \citenamefont
  {Hsu}(2021)}]{wang_simple_2021}%
  \BibitemOpen
  \bibfield  {author} {\bibinfo {author} {\bibfnamefont {S.}~\bibnamefont
  {Wang}}, \bibinfo {author} {\bibfnamefont {Y.-T.}\ \bibnamefont {Chuang}},\
  and\ \bibinfo {author} {\bibfnamefont {L.-Y.}\ \bibnamefont {Hsu}},\
  }\bibfield  {title} {\enquote {\bibinfo {title} {Simple but accurate
  estimation of light–matter coupling strength and optical loss for a
  molecular emitter coupled with photonic modes},}\ }\href
  {https://doi.org/10.1063/5.0060171} {\bibfield  {journal} {\bibinfo
  {journal} {The Journal of Chemical Physics}\ }\textbf {\bibinfo {volume}
  {155}},\ \bibinfo {pages} {134117} (\bibinfo {year} {2021})},\ \bibinfo
  {note} {publisher: American Institute of Physics}\BibitemShut {NoStop}%
\bibitem [{\citenamefont {Herrera}\ and\ \citenamefont
  {Litinskaya}(2022)}]{herrera2022disordered}%
  \BibitemOpen
  \bibfield  {author} {\bibinfo {author} {\bibfnamefont {F.}~\bibnamefont
  {Herrera}}\ and\ \bibinfo {author} {\bibfnamefont {M.}~\bibnamefont
  {Litinskaya}},\ }\bibfield  {title} {\enquote {\bibinfo {title} {Disordered
  ensembles of strongly coupled single-molecule plasmonic picocavities as
  nonlinear optical metamaterials},}\ }\href@noop {} {\bibfield  {journal}
  {\bibinfo  {journal} {The Journal of Chemical Physics}\ }\textbf {\bibinfo
  {volume} {156}},\ \bibinfo {pages} {114702} (\bibinfo {year}
  {2022})}\BibitemShut {NoStop}%
\bibitem [{\citenamefont {Kavokin}\ \emph {et~al.}(2017)\citenamefont
  {Kavokin}, \citenamefont {Baumberg}, \citenamefont {Malpuech},\ and\
  \citenamefont {Laussy}}]{kavokin2017microcavities}%
  \BibitemOpen
  \bibfield  {author} {\bibinfo {author} {\bibfnamefont {A.~V.}\ \bibnamefont
  {Kavokin}}, \bibinfo {author} {\bibfnamefont {J.~J.}\ \bibnamefont
  {Baumberg}}, \bibinfo {author} {\bibfnamefont {G.}~\bibnamefont {Malpuech}},\
  and\ \bibinfo {author} {\bibfnamefont {F.~P.}\ \bibnamefont {Laussy}},\
  }\href@noop {} {\emph {\bibinfo {title} {Microcavities}}},\ Vol.~\bibinfo
  {volume} {21}\ (\bibinfo  {publisher} {Oxford university press},\ \bibinfo
  {year} {2017})\BibitemShut {NoStop}%
\bibitem [{\citenamefont {Akram}, \citenamefont {Ficek},\ and\ \citenamefont
  {Swain}(2000)}]{akram2000decoherence}%
  \BibitemOpen
  \bibfield  {author} {\bibinfo {author} {\bibfnamefont {U.}~\bibnamefont
  {Akram}}, \bibinfo {author} {\bibfnamefont {Z.}~\bibnamefont {Ficek}},\ and\
  \bibinfo {author} {\bibfnamefont {S.}~\bibnamefont {Swain}},\ }\bibfield
  {title} {\enquote {\bibinfo {title} {Decoherence and coherent population
  transfer between two coupled systems},}\ }\href@noop {} {\bibfield  {journal}
  {\bibinfo  {journal} {Physical Review A}\ }\textbf {\bibinfo {volume} {62}},\
  \bibinfo {pages} {013413} (\bibinfo {year} {2000})}\BibitemShut {NoStop}%
\bibitem [{Note1()}]{Note1}%
  \BibitemOpen
  \bibinfo {note} {The choice of intermolecular coupling is $0.1$ in unit of
  Rabi splitting. Rabi splitting is approximately on the order of $10 meV$, and
  if the distance between dipoles are $2 nm$, the dipole moments are on the
  order of $10 Debye$.}\BibitemShut {Stop}%
\bibitem [{Note2()}]{Note2}%
  \BibitemOpen
  \bibinfo {note} {This formula is by modifying equation (7.104) in ref.
  \protect \citenum {nitzan2006chemical} by taking both decay rates $N\Gamma
  _{rad}+\Gamma _{loc}$ into account according to Voigt theorem.}\BibitemShut
  {Stop}%
\bibitem [{\citenamefont {Houdr{\'e}}, \citenamefont {Stanley},\ and\
  \citenamefont {Ilegems}(1996)}]{houdre1996vacuum}%
  \BibitemOpen
  \bibfield  {author} {\bibinfo {author} {\bibfnamefont {R.}~\bibnamefont
  {Houdr{\'e}}}, \bibinfo {author} {\bibfnamefont {R.}~\bibnamefont
  {Stanley}},\ and\ \bibinfo {author} {\bibfnamefont {M.}~\bibnamefont
  {Ilegems}},\ }\bibfield  {title} {\enquote {\bibinfo {title} {Vacuum-field
  rabi splitting in the presence of inhomogeneous broadening: Resolution of a
  homogeneous linewidth in an inhomogeneously broadened system},}\ }\href@noop
  {} {\bibfield  {journal} {\bibinfo  {journal} {Physical Review A}\ }\textbf
  {\bibinfo {volume} {53}},\ \bibinfo {pages} {2711} (\bibinfo {year}
  {1996})}\BibitemShut {NoStop}%
\bibitem [{\citenamefont {Cui}\ and\ \citenamefont
  {Nizan}(2022)}]{cui2022collective}%
  \BibitemOpen
  \bibfield  {author} {\bibinfo {author} {\bibfnamefont {B.}~\bibnamefont
  {Cui}}\ and\ \bibinfo {author} {\bibfnamefont {A.}~\bibnamefont {Nizan}},\
  }\bibfield  {title} {\enquote {\bibinfo {title} {Collective response in
  light--matter interactions: The interplay between strong coupling and local
  dynamics},}\ }\href@noop {} {\bibfield  {journal} {\bibinfo  {journal} {The
  Journal of Chemical Physics}\ }\textbf {\bibinfo {volume} {157}},\ \bibinfo
  {pages} {114108} (\bibinfo {year} {2022})}\BibitemShut {NoStop}%
\bibitem [{\citenamefont {Rybicki}(1994)}]{Rybicki1994notes}%
  \BibitemOpen
  \bibfield  {author} {\bibinfo {author} {\bibfnamefont {G.~B.}\ \bibnamefont
  {Rybicki}},\ }\href@noop {} {\enquote {\bibinfo {title} {Notes on gaussian
  random functions with exponential correlation functions (ornstein-uhlenbeck
  process)},}\ }\bibinfo {howpublished}
  {\url{https://www.lanl.gov/DLDSTP/fast/OU_process.pdf}} (\bibinfo {year}
  {1994})\BibitemShut {NoStop}%
\bibitem [{\citenamefont {P{\'e}rez-S{\'a}nchez}\ \emph
  {et~al.}(2022)\citenamefont {P{\'e}rez-S{\'a}nchez}, \citenamefont {Koner},
  \citenamefont {Stern},\ and\ \citenamefont {Yuen-Zhou}}]{perez2022effective}%
  \BibitemOpen
  \bibfield  {author} {\bibinfo {author} {\bibfnamefont {J.~B.}\ \bibnamefont
  {P{\'e}rez-S{\'a}nchez}}, \bibinfo {author} {\bibfnamefont {A.}~\bibnamefont
  {Koner}}, \bibinfo {author} {\bibfnamefont {N.~P.}\ \bibnamefont {Stern}},\
  and\ \bibinfo {author} {\bibfnamefont {J.}~\bibnamefont {Yuen-Zhou}},\
  }\bibfield  {title} {\enquote {\bibinfo {title} {An effective single molecule
  model that simulates dynamics of collective strong light-matter coupling},}\
  }\href@noop {} {\bibfield  {journal} {\bibinfo  {journal} {arXiv preprint
  arXiv:2209.04955}\ } (\bibinfo {year} {2022})}\BibitemShut {NoStop}%
\bibitem [{\citenamefont {Nitzan}(2006)}]{nitzan2006chemical}%
  \BibitemOpen
  \bibfield  {author} {\bibinfo {author} {\bibfnamefont {A.}~\bibnamefont
  {Nitzan}},\ }\href@noop {} {\emph {\bibinfo {title} {Chemical dynamics in
  condensed phases: relaxation, transfer and reactions in condensed molecular
  systems}}}\ (\bibinfo  {publisher} {Oxford university press},\ \bibinfo
  {year} {2006})\BibitemShut {NoStop}%
\end{thebibliography}%

\end{document}